\documentclass[twocolumn,showpacs,preprintnumbers,amsmath,amssymb,prb,
superscriptaddress]{revtex4}

\usepackage{bm}

\begin{document}


\title{QED of excitons with microscopic nonlocality 
in arbitrary-structured dielectrics}

\author{Motoaki Bamba}
\email{bamba@pe.osakafu-u.ac.jp}
\affiliation{%
Department of Materials Engineering Science,
Graduate School of Engineering Science, Osaka University,
Toyonaka, Osaka 560-8531, Japan
}%
\affiliation{
CREST, Japan Science and Technology Agency, Kawaguchi, Saitama 332-0012, Japan
}%
\author{Hajime Ishihara}
\affiliation{
CREST, Japan Science and Technology Agency, Kawaguchi, Saitama 332-0012, Japan
}%
\affiliation{
Department of Physics and Electronics, Graduate School of Engineering,
Osaka Prefecture University,
Sakai, Osaka 599-8531, Japan
}%

\date{\today}

\begin{abstract}
We have constructed complete quantum theory for an optical process of excitons
with microscopic nonlocality (nonlocal susceptibility)
originating from their center-of-mass motion.
This theory provides a practical calculation method for
arbitrary-structured nano-to-macro scale dielectrics
where excitons are weakly confined.
We obtain good correspondences with underlying theories,
semiclassical microscopic nonlocal theory, and 
QED theories for dispersive and absorptive materials with local susceptibility.
\end{abstract}

\maketitle

\def\dd{{\rm d}}
\def\ee{{\rm e}}
\def\ii{{\rm i}}
\def\vnabla{\mathbf{\nabla}}
\def\grad{\vnabla}
\def\div{\vnabla\cdot}
\def\rot{\vnabla\times}
\def\Lapl{\vnabla^2}
\def\ddt#1{\frac{\partial #1}{\partial t}}
\def\ddtt#1{\frac{\partial^2 #1}{\partial t^2}}
\def\ddz#1{\frac{\partial #1}{\partial z}}
\def\ddzz#1{\frac{\partial^2 #1}{\partial z^2}}
\def\dgg#1{\{#1\}^{\dagger}}
\def\cjg#1{\{#1\}^{*}}
\def\cjgt#1{\{#1\}^{\text{t}*}}
\def\trans#1{\{#1\}^{\text{t}}}
\def\ket#1{|#1\rangle}
\def\bra#1{\langle #1|}
\def\braket#1{\langle #1\rangle}
\def\Ket#1{\left|#1\right\rangle}
\def\Bra#1{\left\langle #1\right|}
\def\Braket#1{\left\langle #1\right\rangle}

\def\const{\text{const.}}
\def\cc{\text{c.c.}}
\def\Hc{\text{H.c.}}
\def\Re{\text{Re}}
\def\Im{\text{Im}}
\def\tr{\text{tr}}
\def\half{\frac{1}{2}}

\def\die{\epsilon}
\def\diez{\epsilon_0}
\def\dieb{\epsilon_{\text{bg}}}
\def\mdiea{\bm{\epsilon}^{\text{abs}}}
\def\muz{\mu_0}
\def\chib{\chi_{\text{bg}}}
\def\chiex{\chi_{\text{ex}}}
\def\mchiex{\bm{\chi}_{\text{ex}}}
\def\mchiexa{\bm{\chi}_{\text{ex}}^{\text{abs}}}

\def\ET{E_{\text{T}}}
\def\EL{E_{\text{L}}}
\def\Eex{E_{\text{ex}}}
\def\wex{\omega}
\def\wres{\Omega}
\def\wT{\omega_{\text{T}}}
\def\DLT{\Delta_{\text{LT}}}
\def\mz{m_0}
\def\mex{m_{\text{ex}}}
\def\mexT{m_{\text{ex}}^{\text{T}}}
\def\mexL{m_{\text{ex}}^{\text{L}}}
\def\damp{\gamma}
\def\dampex{\gamma_{\text{ex}}}
\def\ccxr{g}

\def\rct{\mathcal{A}}
\def\mrct{\bm{\mathsf{A}}}
\def\css{S}
\def\mcss{\bm{\mathsf{S}}}
\def\mS{\bm{\mathsf{S}}}
\def\cssa{S^{\text{abs}}}
\def\mcssa{\bm{\mathsf{S}}^{\text{abs}}}
\def\crd{W}
\def\crda{W^{\text{abs}}}
\def\mcrd{\bm{\mathsf{W}}}
\def\mcrda{\bm{\mathsf{W}}^{\text{abs}}}

\def\wvfex{F}
\def\rmfex{\Phi}
\def\cmfex{G}

\def\vzero{\mathbf{0}}
\def\vunit{\mathbf{e}}
\def\vr{\mathbf{r}}
\def\vp{\mathbf{p}}
\def\vs{\mathbf{s}}
\def\vk{\mathbf{k}}
\def\vR{\mathbf{R}}
\def\vRz{\mathbf{R}_0}
\def\vP{\mathbf{P}}
\def\vPex{\mathbf{P}_{\text{ex}}}
\def\vPbg{\mathbf{P}_{\text{bg}}}
\def\vE{\mathbf{E}}

\def\vdimI{\bm{\mathcal{I}}}
\def\dimP{\mathcal{P}}
\def\vdimP{\bm{\mathcal{P}}}
\def\vdimPT{\bm{\mathcal{P}}^{\text{T}}}
\def\vdimPL{\bm{\mathcal{P}}^{\text{L}}}
\def\vdimE{\bm{\mathcal{E}}}
\def\vdimF{\bm{\mathcal{F}}}

\def\munit{\bm{\mathsf{1}}}
\def\mzero{\bm{\mathsf{0}}}
\def\mdeltaT{\bm{\delta}_{\text{T}}}
\def\mdeltaL{\bm{\delta}_{\text{L}}}
\def\mG{\bm{\mathsf{G}}}
\def\mGren{\bm{\mathsf{G}}_{\text{ren}}}
\def\mGrena{\bm{\mathsf{G}}_{\text{ren}}^{\text{abs}}}

\def\oH{H}
\def\oHem{H_{\text{em}}}
\def\oHint{H_{\text{int}}}
\def\oHmat{H_{\text{mat}}}
\def\oHex{H_{\text{ex}}}
\def\oHeh{H_{\text{eh}}}

\def\oel{\alpha}
\def\oeld{\alpha^{\dagger}}
\def\ohl{\beta}
\def\ohld{\beta^{\dagger}}
\def\oex{b}
\def\oexd{b^{\dagger}}
\def\ores{d}
\def\oresd{d^{\dagger}}
\def\ovA{\mathbf{A}}
\def\ovE{\mathbf{E}}
\def\ophi{\phi}
\def\ophibg{\phi_{\text{bg}}}
\def\ophiex{\phi_{\text{ex}}}
\def\orhobg{\rho_{\text{bg}}}
\def\orhoex{\rho_{\text{ex}}}
\def\oNex{N_{\text{ex}}}
\def\ovIex{\mathbf{I}_{\text{ex}}}
\def\ovJex{\mathbf{J}_{\text{ex}}}
\def\ovPex{\mathbf{P}_{\text{ex}}}

\def\hex{\hat{b}}
\def\hexd{\hat{b}^{\dagger}}
\def\hexz{\hat{b}^{(0)}}
\def\hexzd{\hat{b}^{(0)\dagger}}
\def\hvPex{\hat{\mathbf{P}}_{\text{ex}}}
\def\hJ{\hat{J}}
\def\hJd{\hat{J}^{\dagger}}
\def\hvJ{\hat{\mathbf{J}}}
\def\hvJd{\hat{\mathbf{J}}^{\dagger}}
\def\hvJz{\hat{\mathbf{J}}_0}
\def\hvJza{\hat{\mathbf{J}}_0^{\text{abs}}}
\def\hvE{\hat{\mathbf{E}}}
\def\hvEz{\hat{\mathbf{E}}_0}
\def\hrsrc{\hat{\mathcal{D}}}
\def\hrsrcd{\hat{\mathcal{D}}^{\dagger}}

\section{\label{sec:intro}Introduction}
In the conventional theories of optical processes in condensed matters, 
light has been mainly treated classically 
regardless of whether the matter systems 
are described in quantum mechanical terms (semiclassical theory) 
or classical ones.
These theories have successfully explained 
a variety of optical phenomena for the classical light 
or the coherent states of photons.
However, there is growing interest in the quantum electrodynamics (QED)
of elementary excitations in condensed matters
in order to discuss optical processes for nonclassical light
such as entangled states, single photons, squeezed states, 
cavity photons, and so on.
The relevant experiments have already been reported, for example,
the entangled-photon generation via biexcitons (excitonic molecules),
\cite{edamatsu04}
triggered single photon generation from bound excitons in a semiconductor,
\cite{strauf02}
and the squeezing of cavity polaritons in semiconductor microcavities.
\cite{karr04}
The quantization of a radiation field has been studied for a long time
not only in a vacuum\cite{cohen-tannoudji89} but also in the medium
characterized by a frequency-independent dielectric constant.
On the other hand, Hopfield has systematically discussed
the eigenstates of exciton-photon systems or exciton-polaritons,
\cite{hopfield58} which have a frequency dependence in a susceptibility 
$\chi(\omega)$ or in a dielectric function $\die(\omega) = 1 + \chi(\omega)$ 
as seen from their dispersion relation $\omega^2\die(\omega) = c^2k^2$.
Although in his treatment, $\chi(\omega)$ included only a real part,
susceptibility is generally represented as a complex function satisfying 
the Kramers-Kronig relations.
In addition, its imaginary part, causing damping effects,
cannot be neglected in the discussion of resonant optical processes 
of elementary excitations in condensed matters.

The quantization of the electromagnetic fields
in such dispersive and absorptive dielectrics
has been systematically carried out
for homogeneous media by Huttner and Barnett (HB),\cite{huttner92}
and for inhomogeneous 3D ones by Suttorp and Wubs (SW).\cite{suttorp04}
In the former scheme, dispersive dielectrics 
are described using the Hopfield polariton model,\cite{hopfield58}
i.e., polarizable harmonic oscillators interacting with a radiation field,
and absorption is considered using a reservoir of oscillators
interacting with the polarizable ones.
The electromagnetic fields are described in terms of the eigen operators
derived from the diagonalization of a Hamiltonian.
In the expression of those fields, 
there exists a complex dielectric function $\die(\omega)$
represented by system parameters with satisfying the Kramers-Kronig relations.
All the quantum mechanical properties of the electromagnetic fields
are characterized by this dielectric function.
The pioneering work of HB
stimulated various theoretical studies
associated with the QED of dispersive and absorptive dielectrics,
for example, the spontaneous decay,\cite{barnett92,koshino96}
input-output relations,\cite{matloob95,savasta96,gruner96aug}
and quantization in amplifying, anisotropic, magnetic, or nonlinear media.
\cite{knoll01}
On the other hand, SW have carried out the quantization
of the electromagnetic fields in arbitrary-structured 3D dielectrics
by using the Laplace-transformatin technique,\cite{wubs01,suttorp04}
which is completely different from the quantization scheme of HB.
Around the same time, the diagonalization of the Hamiltonian of SW
has been performed by Suttorp and van Wonderen.\cite{suttorp04EPL}
In these schemes, complex dielectric function $\die(\vr,\omega)$
depends on spatial position $\vr$ of a medium
and radiation frequency $\omega$.

In the above QED theories and also in semiclassical ones,
a dielectric function is usually treated as local form
$\die(\vr,\omega)$ with respect to the spatial position.
However, in general, the optical susceptibility has a nonlocal form as
$\chi(\vr,\vr',\omega)$, which characterizes polarization
$\vP(\vr,\omega)$ at position $\vr$ induced by electric field
$\vE(\vr',\omega)$ at different position $\vr'$ as
\begin{equation} \label{eq:P=chi*E-nonlocal} 
\vP(\vr,\omega) = \diez \int\dd\vr'\ \chi(\vr,\vr',\omega) \vE(\vr',\omega).
\end{equation}
This microscopic nonlocality originates from the spatial spreading of 
the wave function of elementary excitations or, 
particularly for excitons in semiconductors,
their center-of-mass motion with a finite translational mass.
Usually, such a nonlocality is not considered
in the discussion of macroscopic materials;
this is because the coherence length of elementary excitations
is generally much shorter than the spatial scale of materials.
Therefore, only the averaged values of physical quantities 
over the coherence volume are reflected in observation,
and the microscopic nonlocal effect is not apparent.
However, in order to discuss the excitons in inhomogeneous media, 
we must suppose that the motion of excitons has considerably long coherence
and microscopic nonlocality becomes important
even for bulk materials, as explained below.

In the case of homogeneous media,
nonlocal susceptibility depends only on
the difference $\vr-\vr'$ of the two positions;
then, Eq.~\eqref{eq:P=chi*E-nonlocal} is rewritten in the reciprocal space as
\begin{equation}
\vP(\vk,\omega) = \diez \chi(\vk,\omega) \vE(\vk,\omega).
\end{equation}
In this way, susceptibility $\chi(\vk,\omega)$ has a wavevector dependence
and the $\omega$-dependence even for homogeneous media
when it has a microscopic nonlocality.
This $\vk$-dependence gives more than one propagating or evanescent modes 
for a single frequency satisfying
\begin{equation} \label{eq:disp-nonlocal} 
\omega^2\die(\vk,\omega) = c^2|\vk|^2.
\end{equation}
Now, we consider a single exciton state with finite translational mass 
$\mex$, transverse exciton energy $\ET$, and longitudinal one $\EL$
at respective band edges.
Since the transverse exciton energy is written as 
$\Eex(\vk) = \ET~+~\hbar^2|\vk|^2/2\mex$ for wavevector $\vk$, 
we can find two propagating polariton modes for $\hbar\omega > \EL$,
and one propagating mode at the polariton band gap $\ET < \hbar\omega < \EL$
in addition to an evanescent mode.
These multiple polariton modes do not appear in
the Hopfield polariton model\cite{hopfield58}
because the excitons were assumed to have infinite translational mass.
As pointed out by Pekar for the first time,\cite{pekar57}
it appears that additional boundary conditions (ABCs) should be introduced 
besides the Maxwell boundary conditions
for the unique connection between the polariton modes inside a material
and the external ones at the interface between two materials. 
This problem is known as the ABC problem;
it arises when the translational symmetry of a system is broken 
due to surfaces or interfaces. 
Since Pekar's work, continued studies have revealed that this problem 
can be resolved by considering the microscopic boundary conditions 
of the excitonic center-of-mass motion at interfaces.
\cite{zeyher72,dandrea82,cho85}
Nowadays, in the semiclassical framework, 
a calculation method independent from the notation of ABCs 
is well known as an ABC-free theory\cite{cho86} 
or a microscopic nonlocal theory.\cite{cho91,cho03}
These theories systematically consider the nonlocality of susceptibility 
$\chi(\vr,\vr',\omega)$, and various linear and nonlinear phenomena 
in inhomogeneous materials have been discussed using them.
In particular, for nano-structured materials, 
where the coherence of the center-of-mass motion of excitons is maintained 
in the entire material (weak confinement regime), 
anomalous size dependence of their optical processes has been elucidated. 
With regard to nanofilms, the nonlocal theory has successfully explained 
their peculiar spectral structures originating from the polariton 
interference.\cite{cho85,cho90,tang95}
Further, with recent developement of fabrication technologies of nano samples, 
various peculiar effects due to long-range coherence 
are appearing through the interplay between the spatial structures 
of electromagnetic and excitonic waves, 
such as the resonant enhancement of a nonlinear response,\cite{ishihara02jul}
the interchnage of quantized states 
due to giant radiative shift,\cite{syouji04}
and the ultrafast radiative decay with femtosecond order.\cite{ichimiya06}

From these theoretical and experimental results
and great interest in the nonclassical states of elementary 
excitations\cite{edamatsu04,strauf02,karr04} as mentioned above,
it is very attractive to discuss them in detail
for the sake of applications to quantum information technologies.
Although some studies introduced microscopic nonlocality into the QED
of dispersive and absorptive dielectrics, 
there remains a problem of how to perform calculations 
in practical applications, as shown in Sec.~\ref{sec:nonlocalQED}.
The principal purpose of this paper is
to construct a QED theory providing a practical calculation method 
for excitons weakly confined in arbitrary-structured 3D dielectrics 
considering radiative and nonradiative relaxations, 
which is necessary to discuss the effects of, for example, 
material interfaces, excitonic confinement in nano structures, 
and nonradiative relaxation processes. 
In this paper, we merge a microscopic nonlocal theory\cite{cho91,cho03}
and the quantization technique of SW.\cite{suttorp04}

We explain SW's QED theory for dielectrics with local susceptibility
in Sec.~\ref{sec:qmaxwell}
and the previously discussed QED theories with microscopic nonlocality
in Sec.~\ref{sec:nonlocalQED}.
We show our Hamiltonian in Sec.~\ref{sec:hamilt},
explain our QED theory in Sec.~\ref{sec:self-consist},
and discuss the nonradiative relaxation of excitons
in Sec.~\ref{sec:absorption}.
Last, we compare our theory with
other QED ones with microscopic nonlocality in Sec.~\ref{sec:discussion}
and summarize the discussion in Sec.~\ref{sec:summary}.
We will explain only the outline of our theory in these sections
and redundant calculations are shown in appendices.
In App.~\ref{app:2ndJP}, the second quantization of excitonic polarization
is discussed.
In App.~\ref{app:hamilt}, we derive the Hamiltonian 
from the microscopic point of view.
In App.~\ref{app:qmaxwell}, we extend the Maxwell wave equation
discussed by SW so that we can consider the excitonic polarization.
In App.~\ref{app:qlinopt}, we evaluate the commutation relations
of $\omega$-Fourier transformed Heisenberg operators.
In App.~\ref{app:ssm}, we discuss the relationship between the retarded
correlation functions of excitons and self-standing modes 
or exciton polariton modes.
In App.~\ref{app:stcr}, we verify the equal-time commutation relations,
which are expected from those of Shr\"{o}dinger operators.

In this paper, we use MKS units and Coulomb gauge.

\section{\label{sec:qmaxwell}QED of dielectrics without nonlocality}
In this section, we explain the outline of the discussion of SW,
\cite{suttorp04}
the QED theory without microscopic nonlocality.
Its Hamiltonian is shown in Eq.~\eqref{eq:def-Hem} of the present paper.
The quantization scheme is based on the motion equation of the electric field
$\ovE(\vr,t)$ whose definition in SW theory is
\begin{equation} \label{eq:def-E-bg} 
\ovE(\vr,t) = - \ddt{}\ovA(\vr,t) - \grad\ophibg(\vr,t).
\end{equation}
$\ovA(\vr,t)$ is the vector potential and 
$\ophibg(\vr,t)$ is the Coulomb potential.
We write the latter and the dielectric function $\dieb(\vr,\omega)$
with a subscript $\text{bg}$ for the description in the following sections.
Since the Coulomb gauge is used in this scheme,
the vector potential is a transverse field satisfying $\div\ovA(\vr) = 0$,
and the second term of Eq.~\eqref{eq:def-E-bg} represents 
the longitudinal field.
From the Lapalce transform of the Heisenberg equations 
of the system variables,
the Maxwell wave equation for the Fourier component of the electric field
\begin{equation} \label{eq:def-hvE} 
\hvE^{\pm}(\vr,\omega) \equiv \frac{1}{2\pi} \int_{-\infty}^{\infty}\dd t\
\ee^{\pm\ii\omega t} \ovE(\vr,t)
= \dgg{\hvE^{\mp}(\vr,\omega)}
\end{equation}
is derived as
\begin{equation} \label{eq:Maxwell-E-bg} 
\rot\rot\hvE^+(\vr,\omega)
- \frac{\omega^2}{c^2}\dieb(\vr,\omega)\hvE^+(\vr,\omega)
= \ii\muz\omega\hvJ(\vr,\omega),
\end{equation}
where $\hvE^+(\vr,\omega)$ and $\hvE^-(\vr,\omega)$ are called
positive- and negative-frequency Fourier components of $\ovE(\vr,t)$, 
respectively.
We write the Fourier transformed operator with a hat ( $\hat{}$ ) 
in this paper.
Wave equation \eqref{eq:Maxwell-E-bg} has the same form
as the one that appears in the classical electrodynamics
except for operator $\hvJ(\vr,\omega)$ on the right hand side.
This operator is called the noise current density and interpreted as
a source of the electromagnetic fields or the fluctuation caused by absorption.
It plays an essential role in the series of QED theories
for dispersive and absorptive dielectrics.
The same kind of operator for homogeneous systems
has been derived by HB,\cite{huttner92}
and the one for inhomogeneous 3D systems
has been phenomenologically introduced in Ref.~\onlinecite{dung98}.
On the other hand, from the Lapalce-transformed motion equations
of system variables,
SW have systematically derived
the representation of $\hvJ(\vr,\omega)$, which is written
in terms of the canonical variables and momenta of the system at $t = 0$
(we have performed a similar calculation in App.~\ref{app:qmaxwell}).
From the commutation relations between them,
those for $\hvJ(\vr,\omega)$ have been derived as
\begin{subequations} \label{eq:[hvJ,hvJ]} 
\begin{align}&
\left[ \hvJ(\vr,\omega), \dgg{\hvJ(\vr',\omega')} \right]
\nonumber \\ & \quad
= \delta(\omega-\omega') \delta(\vr-\vr') \frac{\diez\hbar\omega^2}{\pi}
    \Im[\dieb(\vr,\omega)] \munit, \\
& \left[ \hvJ(\vr,\omega), \hvJ(\vr',\omega') \right] = \mzero,
\end{align}
\end{subequations}
where $[ \hvJ, \hvJd ]$ is a $3\times3$ tensor 
and its $(\xi,\xi')$ element implies $[\hJ_{\xi},\hJd_{\xi'}]$
for $\xi = x, y, z$
(the same kind of calculation is shown in App.~\ref{app:qlinopt}
of the present paper).

Here, using Green's tensor $\mG(\vr,\vr',\omega)$ satisfying
\begin{equation} \label{eq:satisfied-mG} 
\rot\rot\mG(\vr,\vr',\omega)
- \frac{\omega^2}{c^2}\dieb(\vr,\omega)\mG(\vr,\vr',\omega)
= \delta(\vr-\vr')\munit,
\end{equation}
we can rewrite Maxwell wave equation \eqref{eq:Maxwell-E-bg} as
\begin{equation} \label{eq:E=GJ-bg} 
\hvE^+(\vr,\omega)
= \ii\muz\omega \int\dd\vr'\ \mG(\vr,\vr',\omega) \cdot \hvJ(\vr',\omega).
\end{equation}
From the commutation relations \eqref{eq:[hvJ,hvJ]} of $\hvJ(\vr,\omega)$,
those of $\hvE^{\pm}(\vr,\omega)$ can be derived as
\begin{subequations} \label{eq:[E(r,w),E(r,w)]-bg} 
\begin{align}
\left[ \hvE^+(\vr,\omega), \hvE^-(\vr',\omega') \right]
& = \delta(\omega-\omega') \frac{\muz\hbar\omega^2}{\pi}
    \Im[ \mG(\vr,\vr',\omega) ], \\
\left[ \hvE^+(\vr,\omega), \hvE^+(\vr',\omega') \right]
& = \mzero,
\end{align}
\end{subequations}
where we use the equivalence shown in Eq.~(1.54) of Ref.~\onlinecite{knoll01}:
\begin{align}& \label{eq:G(e-e)G=G-G} 
\int\dd\vs\ \frac{\omega^2}{c^2} \Im[\dieb(\vs,\omega)]
   \mG(\vr,\vs,\omega) \cdot \mG^*(\vs,\vr',\omega)
\nonumber \\ &
= \Im[\mG(\vr,\vr',\omega)]
\end{align}
and the reciplocity relation:
\begin{equation} \label{eq:recip-mG} 
\mG(\vr,\vr',\omega) = \trans{\mG(\vr',\vr,\omega)}.
\end{equation}
Eqs.~\eqref{eq:[E(r,w),E(r,w)]-bg} can be understood by the fact that
the Green's tensor $\mG(\vr,\vr',\omega)$ for the Maxwell wave equation
\eqref{eq:Maxwell-E-bg} identifies with the Fourier transform of 
the retarded correlation function of the electric field
except for the factor $-\muz\hbar\omega^2$,
as discussed in Ref.~\onlinecite{abrikosov75ch6}:
\begin{align}&
- \muz\hbar\omega^2\mG(\vr,\vr',\omega)
\nonumber \\ &
= \int_{-\infty}^{\infty}\dd t\ \ee^{\ii\omega(t-t')}
  (-\ii) \theta(t-t') \braket{[\vE(\vr,t), \vE(\vr',t')]},
\end{align}
where $\theta(t)$ is the Heaviside step function.
In this framework, 
all we have to do is find Green's tensor satisfying 
Eq.~\eqref{eq:satisfied-mG}
in order to discuss the quantum mechanical properties of 
the electromagnetic fields in dielectrics.
All the information of the material structure is included
in dielectric functin $\dieb(\vr,\omega)$.
The form of Green's tensor has already been known 
for various structures with high symmetry\cite{chew95}
and also can be numerically calculated 
for arbitrary 3D structures.\cite{martin95}

\section{\label{sec:nonlocalQED}Previous QED theories with microscopic 
nonlocality}
The above theory enables us to discuss the linear optical process
in arbitrary-structured 3D dielectrics characterized
by dielectric function $\dieb(\vr,\omega)$.
However, 
in order to discuss the materials with microscopic nonlocality,
it is necessary to consider more general elementary excitations
that cannot be described by harmonic oscillators of the Hopfield model.

As a pioneering study on a full-quantum theory with microscopic nonlocality, 
Jenkins and Mukamel have discussed molecular crystals in $d$ dimensions
($d = 1, 2, 3$),\cite{jenkins93}
where the relative motion of excitons is localized at a single molecule
and the center-of-mass moves between molecules 
due to dipole-dipole interaction.
While their theory concentrates on treating the resonant polarization
without nonradiative relaxation,
recently, microscopic nonlocality is being introduced into 
the field quantization in dispersive and absorptive media,
\cite{stefano99,stefano01,savasta02pra,bechler06,suttorp07,raabe07}
and some studies have demonstrated the application of their theories 
for specific structures.\cite{stefano99,raabe07}
Stefano et al.~discussed excitons with the microscopic nonlocality
in media with spatial translation symmetry broken along one dimension,
and they practically calculated the spatial and frequency dependences 
of the vacuum-field flctuation in a semiconductor quantum well structure.
\cite{stefano99}
Thereafter, they extended their theory to an arbitrary 3D structure,
\cite{stefano01} and discussed the input-output relations 
in scattering systems.\cite{savasta02pra}
On the other hand, 
Bechler performed the field quantization for the homogeneous systems
with nonlocality by using the path-integral method,\cite{bechler06}
and Suttorp did for the nonlocal, inhomogeneous, and anisotropic systems
by using the diagnalization method.\cite{suttorp07}
Most recently, Raabe et al.~phenomenologically discussed
the nonlocal systems with both dielectric and magnetic properties.
\cite{raabe07}
They propose the use of the dielectric approximation with the surface 
impedance method for the practical application of their theory.

As seen in the above studies, 
it is safe to say that a consistent framework for the field quantization 
in dielectrics with microscopic nonlocality
has already been established.
Thus, the issue of current importance is to establish a general and 
practical calculation method applicable to arbitrary-structured 3D system, 
which is desired for the actual applications of the above framework 
though interesting applications have already been demonstrated 
in specific situations by Stefano et al.~and Raabe et al.
The essential task for this purpose is the derivation of Green's tensor
for the Maxwell wave equation with nonlocal susceptibility
as seen in Eq.~\eqref{eq:Maxwell-nonlocal-damp} of the present paper.

In this paper, we provide a practical 
calculation method for Green's tensor for arbitrary structures
by using the fact that the nonlocal susceptibility has a separable form
with respect to two positions as seen in Eq.~\eqref{eq:def-mchiexa}.
This technique has been developed in the semiclassical microscopic nonlocal 
theory.\cite{cho91,cho03}
We extend this theory
to be able to consider the quantum mechanical properties of
electromagnetic fields
by using the Laplace transformation technique of SW.\cite{suttorp04}
In other words, we generalize the SW theory to media with
microscopic nonlocality for providing the practical calculation method.
The SW theory is suitable to extend the nonlocal theory
because the latter consists of two equatons:
the Maxwell wave equation,
which is just the fundamental one in the SW theory as seen above,
and the motion equation of excitonic polarization,
which can easily be derived by the Laplace transformation technique.

We provide the outline of our theory in the following sections, 
and the detailed explanations, 
including lengthy calculations, are given in Apps.~\ref{app:2ndJP} 
to \ref{app:qlinopt} to keep strightforwardness of the main part.
After that, we will compare our theory 
with some of the above-mentioned previously discussed ones.

\section{\label{sec:hamilt}Hamiltonian}
We describe the dielectric materials 
with resonant contributions from excitons with center-of-mass motion 
and the nonresonant ones with local dielectric function
$\dieb(\vr,\omega)$.
This treatment is essential for including the consideration of the effects 
arising from the radiation mode structures modified by the practical dielectric
structures (with absorption) such as a substrate, 
a dielectric multilayer cavity, photonic crystals, and so on, 
surrounding excitonic active structures.    
We explicitly discuss optical and nonradiative-relaxation processes 
of the excitons, and the nonresonant backgrounds are treated 
as the same procedure of SW.\cite{suttorp04}
The total Hamiltonian discussed in the present paper is written as
\begin{equation} \label{eq:rep-oH} 
\oH = \oHem + \oHint + \oHmat,
\end{equation}
where $\oHem$ describes the radiation field and background dielectric medium,
$\oHmat$ represents excitons and a reservoir of oscillators,
and $\oHint$ is the interactions between $\oHem$ and $\oHmat$.
$\oHem$ is just the complete Hamiltonian discussed
by SW\cite{suttorp04},
and its representation is shown in Eq.~\eqref{eq:def-Hem} of the present paper.

As a result of the discussion in App.~\ref{app:hamilt},
the interaction Hamlitonian is represented as
\begin{align} \label{eq:rep-oHint} 
\oHint
& = - \int\dd\vr \left[ \ovIex(\vr) \cdot \ovA(\vr)
                      - \frac{1}{2} \oNex(\vr)\ovA^2(\vr) \right]
\nonumber \\ & \quad
  + \int\dd\vr\ \ophibg(\vr) \orhoex(\vr)
  + \half \int\dd\vr\ \ophiex(\vr) \orhoex(\vr).
\end{align}
$\ovIex(\vr)$ is the excitonic current density 
without radiation contribution $ - \oNex(\vr)\ovA(\vr)$, 
i.e., the complete current density is written as 
$\ovJex(\vr) = \ovIex(\vr) - \oNex(\vr)\ovA(\vr)$
(see App.~\ref{app:2ndJP}).
$\orhoex(\vr)$ is the excitonic charge density and
\begin{equation} \label{eq:def-ophiex} 
\ophiex(\vr) \equiv \int\dd\vr'\ \frac{\orhoex(\vr')}{4\pi\diez|\vr-\vr'|}
\end{equation}
is the Coulomb potential.
The first and second terms of Eq.~\eqref{eq:rep-oHint} represent
the interaction between the radiation field and excitons.
The third term is the Coulomb interaction between
induced charges of excitons and that of the background medium.
The last term is the one between excitonic charges themselves,
and is also considered as the dipole-dipole interaction
between excitonic polarizations,
or the exchange interaction between electrons and holes
\cite{cho99,ajiki00,cho03} (see App.~\ref{app:hamilt}).
Although this term usually belongs to matter Hamiltonian $\oHmat$,
we displace it into $\oHint$ because it can also be considered
as the interaction between the longitudinal component of the polarization
and that of the electric field.
This treatment will give us the motion equation of excitons as
a simple form as Eq.~\eqref{eq:motion-hexone},
and will take away our explicit consideration of the longitudinal-transverse 
(LT) splitting of the exciton eigenenergies,
because the last term of Eq.~\eqref{eq:rep-oHint} is the origin of
the LT splitting.

With regard to excitons, generally, 
we should describe them starting from the basis of electrons and holes 
interacting with each other and themselves.
However, as long as we consider a linear optical process of exciotns
under weak excitation,
it is valid to describe electronic systems 
in terms of excitonic eigenstates.
In addition, in order to describe a nonradiative relaxation process,
we consider a reservoir of oscillators interacting with excitons.
The matter Hamiltonian is written as
\begin{align}
\oHmat & = \sum_{\mu} \hbar\wex_{\mu} \oexd_{\mu} \oex_{\mu}
+ \sum_{\mu} \int_0^{\infty}\dd\wres\ \bigl\{
    \hbar\wres\ \oresd_{\mu}(\wres) \ores_{\mu}(\wres)
\nonumber \\ & \quad
+ \left[ \oex_{\mu} + \oexd_{\mu} \right]
  \left[ \ccxr_{\mu}(\wres) \ores_{\mu}(\wres)
       + \ccxr^*_{\mu}(\wres) \oresd_{\mu}(\wres) \right] \bigr\},
\label{eq:rep-oHmat} 
\end{align}
where $\oex_{\mu}$ is the annihilation operator of the excitons
in eigenstate $\mu$ with eigenfrequency $\wex_{\mu}$, 
which does not include the LT splitting 
because we displace the exchange interaction 
between electrons and holes from $\oHmat$ to $\oHint$.
In this paper, we assume that the center-of-mass motion of excitons 
is confined 
in finite spaces, and index $\mu$ represents degrees of freedom 
of not only the relative motion but also the translational one.
Instead of evaluating the commutation relations of $\oex_{\mu}$
from its representation \eqref{eq:def-ex-opr}
with Fermi's commutation relations of electrons and holes,
we consider the excitons as pure bosons satisfying
\begin{subequations} \label{eq:[oex,oex]} 
\begin{align}
[ \oex_{\mu}, \oexd_{\mu'} ] & = \delta_{\mu,\mu'}, \\
[ \oex_{\mu}, \oex_{\mu'} ] & = 0.
\end{align}
\end{subequations}
This approximation is valid under weak excitation.
On the other hand, in Eq.~\eqref{eq:rep-oHmat},
$\ores_{\mu}(\wres)$ is the annihilation operator of the reservoir 
oscillators with frequency $\wres$
interacting with the excitons in state $\mu$,
and $\ccxr_{\mu}(\wres)$ is its coupling parameter.
The oscillators are independent of each other
and satisfy the commutation relations as
\begin{subequations} \label{eq:[ores,ores]} 
\begin{align}
[ \ores_{\mu}(\wres), \oresd_{\mu'}(\wres') ]
& = \delta_{\mu,\mu'} \delta(\wres-\wres'), \\
[ \ores_{\mu}(\wres), \ores_{\mu'}(\wres') ]
& = 0.
\end{align}
\end{subequations}

\section{\label{sec:self-consist}QED of excitons}
In this section,
from motion equations of excitons and 
the Maxwell wave equation for our system derived by the quantization technique 
of SW,\cite{suttorp04}
we show our QED theory for excitons by using the 
technique of the microscopic nonlocal theory developed
in the semiclassical framework.\cite{cho91,cho03}
Since we require a lengthy calculation based on the SW's scheme
to prove the validity of the following discussion,
we show only the outline of our theory and the details are discussed
in Apps.~\ref{app:qmaxwell} and \ref{app:qlinopt}.
We do not consider the nonradiative relaxation of excitons in this section.
It will be discussed in the next section.

Since we consider excitons and also the background medium,
the exciton-induced longitudinal field also contributes 
to the electric field as
\begin{equation} \label{eq:rep-ovE} 
\ovE(\vr,t) = - \ddt{}\ovA(\vr,t) - \grad\ophibg(\vr,t) - \grad\ophiex(\vr,t).
\end{equation}
This definition is different from that of SW
(Eq.~\eqref{eq:def-E-bg} of the present paper).
In addition, instread of Eq.~\eqref{eq:Maxwell-E-bg},
excitonic polarization density $\hvPex^+(\vr,\omega)$
appears in the Maxwell wave equation as
\begin{align}&
\rot\rot\hvE^+(\vr,\omega)
- \frac{\omega^2}{c^2}\dieb(\vr,\omega)\hvE^+(\vr,\omega)
\nonumber \\ &
= \ii\muz\omega\hvJz(\vr,\omega) + \muz\omega^2\hvPex^+(\vr,\omega)
\label{eq:Maxwell-E-Jz-Pex} 
\end{align}
(see App.~\ref{app:qmaxwell}),
where $\hvJz(\vr,\omega)$ is the same kind of operator as
$\hvJ(\vr,\omega)$ in Eq.~\eqref{eq:Maxwell-E-bg},
and it satisfies the same commutation relations of Eq.~\eqref{eq:[hvJ,hvJ]}:
\begin{subequations} \label{eq:[hvJz,hvJz]} 
\begin{align}
&   \left[ \hvJz(\vr,\omega), \dgg{\hvJz(\vr',\omega')} \right]
\nonumber \\ & \quad
= \delta(\omega-\omega') \delta(\vr-\vr') \frac{\diez\hbar\omega^2}{\pi}
    \Im[\dieb(\vr,\omega)] \munit
\label{eq:[hvJz,hvJzd]} \\ 
& \left[ \hvJz(\vr,\omega), \hvJz(\vr',\omega') \right] = \mzero,
\label{eq:[hvJzd,hvJzd]} 
\end{align}
\end{subequations}
(see App.~\ref{app:qlinopt}).
This operator is interpreted as the source of 
the background electromagnetic fields
and its definition is shown in Eq.~\eqref{eq:def-Jz}.
Using Green's tensor $\mG(\vr,\vr',\omega)$
satisfying Eq.~\eqref{eq:satisfied-mG},
we can rewrite Eq.~\eqref{eq:Maxwell-E-Jz-Pex} as
\begin{equation} \label{eq:E=Ez+G*Pex} 
\hvE^+(\vr,\omega) = \hvEz^+(\vr,\omega)
  + \muz\omega^2 \int\dd\vr'\ \mG(\vr,\vr',\omega) \cdot \hvPex^+(\vr',\omega),
\end{equation}
where
\begin{equation} \label{eq:def-hvEz-hvJz} 
\hvEz^+(\vr,\omega) \equiv \ii\muz\omega
  \int\dd\vr'\ \mG(\vr,\vr',\omega) \cdot \hvJz(\vr',\omega)
\end{equation}
is the background electric field satisfying the commutation relations
\begin{subequations} \label{eq:[Ez(r,w),Ez(r,w)]} 
\begin{align}
\left[ \hvEz^+(\vr,\omega), \hvEz^-(\vr',\omega') \right]
& = \delta(\omega-\omega') \frac{\muz\hbar\omega^2}{\pi}
    \Im[\mG(\vr,\vr',\omega)], \label{eq:[hvEz+,hvEz-]} \\ 
\left[ \hvEz^+(\vr,\omega), \hvEz^+(\vr',\omega') \right]
& = \mzero. \label{eq:[hvEz+,hvEz+]} 
\end{align}
\end{subequations}
These are equivalent to relations \eqref{eq:[E(r,w),E(r,w)]-bg} 
of the electric field in a local system.
This equivalence indicates that
the behavior of the background field in our system
is exactly the same as that of the complete electric field
in the local system.
This natural result can be systematically derived
from the representation of $\hvJz(\vr,\omega)$ \eqref{eq:def-Jz}
and commutation relations of the system variables
by using the Laplace transfromation technique.
We have verified this equivalence for the linear optical process of excitons
even with the nonradiative relaxation of excitons (see App.~\ref{app:qlinopt}).

Next, we discuss the motion of unknown variable $\hvPex^+(\vr,\omega)$
in the Maxwell wave equation \eqref{eq:Maxwell-E-Jz-Pex}.
The second-quantized polarization density is written in terms of 
exciton operator set $\{\oex_{\mu}\}$ as
\begin{equation} \label{eq:rep-ovPex-oex} 
\ovPex(\vr)
= \sum_{\mu} \left[ \vdimP_{\mu}(\vr)\ \oex_{\mu}
                  + \vdimP_{\mu}^*(\vr)\ \oexd_{\mu} \right],
\end{equation}
where the expansion coefficient $\vdimP_{\mu}(\vr)$ is
\begin{equation} \label{eq:rep-vdimP} 
\vdimP_{\mu}(\vr) = \dimP_{\mu} \vunit_{\mu} \cmfex_{\mu}(\vr)
\end{equation}
(see App.~\ref{app:2ndJP}).
$\dimP_{\mu}$ is the transition dipole moment,
$\vunit_{\mu}$ is a unit vector in the polarization direction,
and $\cmfex_{\mu}(\vr)$ is the wave function of 
the center-of-mass motion in exciton state $\mu$.
Since we assume the weak confinement regime, 
$\dimP_{\mu}$ approximately depends only on the relative motion of excitons
and is related with LT splitting 
$\DLT^{\mu} = |\dimP_{\mu}|^2/\diez\dieb$.
Neglecting the reservoir oscillators and assumimg $\omega \sim \wex_{\mu}$,
the Fourier-transformed Heisenberg equation of excitons is derived
from Hamiltonian \eqref{eq:rep-oHint} and \eqref{eq:rep-oHmat} as
\begin{equation} \label{eq:motion-hexone} 
(\hbar\wex_{\mu}-\hbar\omega-\ii\delta)\ \hex_{\mu}(\omega)
= \int\dd\vr\ \vdimP_{\mu}^*(\vr) \cdot \hvE^+(\vr,\omega)
\end{equation}
(see App.~\ref{sec:eq-lin-opt-ex}),
where infinitesimal damping $\ii\delta$ is added to match the discussion
with the nonradiative relaxation in the next section.
Under the rotating wave approximation (RWA),
the positive-frequency Fourier component of the polarization density
\eqref{eq:rep-ovPex-oex} is written as
\begin{equation} \label{eq:Pex=P*ex-rwa} 
\hvPex^+(\vr,\omega)
= \sum_{\mu} \vdimP_{\mu}(\vr) \hex_{\mu}(\omega).
\end{equation}
Substituting Eq.~\eqref{eq:motion-hexone} into \eqref{eq:Pex=P*ex-rwa},
we obtain the nonlocal form of the polarization density as
\begin{equation} \label{eq:Pexone=chi*E} 
\hvPex^+(\vr,\omega)
= \diez\int\dd\vr'\ \mchiex(\vr,\vr',\omega) \cdot \hvE^+(\vr',\omega),
\end{equation}
where the susceptibility tensor is defined as
\begin{equation} \label{eq:def-chi-ex} 
\mchiex(\vr,\vr',\omega) \equiv \frac{1}{\diez} \sum_{\mu}
  \frac{\vdimP_{\mu}(\vr)\vdimP_{\mu}^*(\vr')}
       {\hbar\wex_{\mu}-\hbar\omega-\ii\delta}.
\end{equation}
The spatial spreading of the exciton state,
the origin of the nonlocality, is reflected 
through polarization coefficient $\vdimP_{\mu}(\vr)$
or center-of-mass wave function $\cmfex_{\mu}(\vr)$.
On the other hand, 
the spatial structure of the background dielectrics is characterized
by dielectric function $\dieb(\vr,\omega)$ in Maxwell wave equation
\eqref{eq:Maxwell-E-Jz-Pex} 
and in commutation relations \eqref{eq:[hvJz,hvJz]}.
In our framework, we can discuss arbitrary-structured exciton motions
and background dielectrics as seen below.

In order to discuss the optical process of excitons,
we must simultaneously solve Maxwell wave equation 
\eqref{eq:Maxwell-E-Jz-Pex}
and the motion equation of polarization density \eqref{eq:Pexone=chi*E}
to determine unknown variables $\hvE^+(\vr,\omega)$
and $\hvPex^+(\vr,\omega)$.
Substituting Eq.~\eqref{eq:Pexone=chi*E} into \eqref{eq:Maxwell-E-Jz-Pex},
we obtain the nonlocal wave equation as
\begin{align}&
\rot\rot\hvE^+(\vr,\omega)
- \frac{\omega^2}{c^2}\dieb(\vr,\omega)\hvE^+(\vr,\omega)
\nonumber \\ & \quad
- \frac{\omega^2}{c^2} \int\dd\vr'\ \mchiex(\vr,\vr',\omega)
  \cdot \hvE^+(\vr',\omega)
= \ii\muz\omega\hvJz(\vr,\omega). \label{eq:Maxwell-nonlocal} 
\end{align}
This is the same equation discussed 
by Savasta et al.\cite{stefano01,savasta02pra} and 
also has the same form as that of Raabe et al.\cite{raabe07}
However, it appears very difficult to solve this nonlocal equation,
and there remains a problem to derive Green's tensor for this equation
in the practical application of their theories.
This problem  can be solved by using the fact 
that nonlocal susceptibility \eqref{eq:def-chi-ex} 
has a separable form with respect to $\vr$ and $\vr'$.
One scheme is to directly derive Green's tensor 
for Eq.~\eqref{eq:Maxwell-nonlocal} as discussed in Ref.~\onlinecite{cho03jl}, 
and the other is to reduce this nonlocal problem 
into a simultaneous linear equation set.\cite{cho91,cho03}
In our QED theory, we adopt the latter scheme
becuase it provides not only Green's tensor of the former
but also much interesting information on exciton-polariton systems.

Substituting the representation of the electric field \eqref{eq:E=Ez+G*Pex}
into the motion equatin of excitons \eqref{eq:motion-hexone}
by expanding $\hvPex^+(\vr,\omega)$ as Eq.~\eqref{eq:Pex=P*ex-rwa},
we obtain the linear equation set
determing exciton amplitudes $\{\hex_{\mu}(\omega)\}$ as
\begin{align}&
\sum_{\mu'} \left[ 
    (\hbar\wex_{\mu} - \hbar\omega - \ii\delta) \delta_{\mu,\mu'}
  + \rct_{\mu,\mu'}(\omega) \right] \hex_{\mu'}(\omega)
\nonumber \\ &
= (\hbar\wex_{\mu}-\hbar\omega-\ii\delta) \hexz_{\mu}(\omega).
 \label{eq:self-consist} 
\end{align}
This has the same form of the self-consistent equation set 
in the semiclassical nonlocal theory.\cite{cho91,cho03}
The operator on the right hand side
\begin{equation} \label{eq:def-hexz} 
\hexz_{\mu}(\omega) \equiv \frac{1}{\hbar\wex_{\mu}-\hbar\omega-\ii\delta}
  \int\dd\vr\ \vdimP_{\mu}^*(\vr) \cdot \hvEz^+(\vr,\omega)
\end{equation}
has the same form of Eq.~\eqref{eq:motion-hexone} replacing 
$\hvE^+(\vr,\omega)$ with $\hvEz^+(\vr,\omega)$;
then, it can be interpreted as the exciton amplitude
directly induced by the background electric field.
Here, we use the word ``directly'' to mean 
that $\hexz_{\mu}(\omega)$ does not include
the diffusion of the exciton amplitudes via the electromagnetic fields.
Such effect is reflected in the correction term
of eigenenergy $\hbar\wex_{\mu}$ in Eq.~\eqref{eq:self-consist}:
\begin{equation} \label{eq:def-rct} 
\rct_{\mu,\mu'}(\omega)
\equiv - \muz\omega^2 \int\dd\vr\int\dd\vr'\
  \vdimP_{\mu}^*(\vr) \cdot \mG(\vr,\vr',\omega) \cdot \vdimP_{\mu'}(\vr').
\end{equation}
This value represents the exciton-exciton interaction 
via the electromagnetic fields, 
i.e., the polarization at $\vr'$ induces electric field,
and later it induces another polarization at $\vr$.
The interaction between the transverse fields is the retarded 
interaction, and the one between the longitudinal fields is interpreted 
as the Coulomb interaction between induced charges.
The latter is just the exchange interaction between electrons and holes,
which we displace from $\oHmat$ to $\oHint$,
and it gives the LT splitting of the exciton eigenenergies.
The radiative relaxation of excitons and 
the transition between exciton states are described
by the imaginary part of $\rct_{\mu,\mu'}(\omega)$.

Writing the coefficient matrix element of Eq.~\eqref{eq:self-consist} as
\begin{equation} \label{eq:def-element-mS} 
S_{\mu,\mu'}(\omega) 
\equiv (\hbar\wex_{\mu}-\hbar\omega-\ii\delta) \delta_{\mu,\mu'}
  + \rct_{\mu,\mu'}(\omega),
\end{equation}
the self-consistent equation set is rewritten as
\begin{equation} \label{eq:self-consist-S} 
\sum_{\mu'} S_{\mu,\mu'}(\omega) \hex_{\mu'}(\omega)
= \int\dd\vr\ \vdimP_{\mu}^*(\vr) \cdot \hvEz^+(\vr,\omega).
\end{equation}
By using the inverse matrix $\mcrd(\omega) = [\mS(\omega)]^{-1}$
with the basis of exciton states,
we obtain the representation of exciton operators as
\begin{equation} \label{eq:hexone=rct*hexz} 
\hex_{\mu}(\omega) = \sum_{\mu'} \crd_{\mu,\mu'}(\omega)
\int\dd\vr\ \vdimP_{\mu'}^*(\vr) \cdot \hvEz^+(\vr,\omega).
\end{equation}
We can describe all the physical variables in terms of these operators
and $\hvEz^{\pm}(\vr,\omega)$.
For example, 
the excitonic polarization is written as Eq.~\eqref{eq:Pex=P*ex-rwa}
and the electric field \eqref{eq:E=Ez+G*Pex} as
\begin{equation} \label{eq:hvEone=hvEz+E*hexone} 
\hvE^+(\vr,\omega)
= \hvEz^+(\vr,\omega)
+ \sum_{\mu} \vdimE_{\mu}(\vr,\omega) \hex_{\mu}(\omega),
\end{equation}
where the coefficients are defined as
\begin{align}
\vdimE_{\mu}(\vr,\omega)
& \equiv \muz\omega^2 \int\dd\vr'\ \mG(\vr,\vr',\omega)
  \cdot \vdimP_{\mu}(\vr'),
\label{eq:def-vdimE} \\ 
\vdimF_{\mu}(\vr,\omega)
& \equiv \muz\omega^2 \int\dd\vr'\ \mG(\vr,\vr',\omega)
  \cdot \vdimP_{\mu}^*(\vr').
\label{eq:def-vdimF} 
\end{align}
The latter will appear in Eq.~\eqref{eq:rep-mGren}.
The exciton operators are represented by the background electric field
$\hvEz^+(\vr,\omega)$, whose commutation relations are described 
by Green's tensor $\mG(\vr,\vr',\omega)$ as shown 
in Eq.~\eqref{eq:[Ez(r,w),Ez(r,w)]}.
Since there is no problem in deriving Green's tensor as explained above,
the additional works are to perform integrations
\eqref{eq:def-rct}, \eqref{eq:def-vdimE}, and \eqref{eq:def-vdimF},
and to derive inverse matrix $\mcrd(\omega) = [\mS(\omega)]^{-1}$.
These calculations can be performed straightforwardly.
Therefore, in our framework, 
for a given wave function of excitons' center-of-mass motion
$\{\cmfex_{\mu}(\vr)\}$ and 
background dielectric function $\dieb(\vr,\omega)$,
we can discuss the QED of excitonic materials with the microscopic nonlocality.

When we do not consider the nonradiative relaxation of excitons,
all the quantum mechanical properties
are described by commutation relations \eqref{eq:[hvJz,hvJz]}
of noise current density $\hvJz(\vr,\omega)$.
Using relations \eqref{eq:[Ez(r,w),Ez(r,w)]} derived 
from Eqs.~\eqref{eq:[hvJz,hvJz]},
the commutation relations of exciton operators \eqref{eq:hexone=rct*hexz}
are evaluated as
\begin{subequations} \label{eq:[hexone,hexone]} 
\begin{align}&
\left[ \hex_{\mu}(\omega), \dgg{\hex_{\mu'}(\omega')} \right]
\nonumber \\ & \quad
= \delta(\omega-\omega')\ \frac{\hbar}{\ii2\pi}
  \left[ \crd_{\mu,\mu'}(\omega) - \crd_{\mu',\mu}^*(\omega) \right], \\
& \left[ \hex_{\mu}(\omega), \hex_{\mu'}(\omega') \right]
= 0.
\end{align}
\end{subequations}
This result indicates that the elements of inverse matrix $\mcrd(\omega)$ 
of the self-consistent equation set identify
with the Fourier transforms of the retarded correlation functions of excitons
except for the factor $-\hbar$ (see App.~\ref{app:ssm}).
In addition, we can obtain the commutation relations 
of electric field operators \eqref{eq:hvEone=hvEz+E*hexone}:
\begin{subequations} \label{eq:[hvEone,hvEone]} 
\begin{align} &
\left[ \hvE^+(\vr,\omega), \hvE^-(\vr',\omega') \right]
= \delta(\omega-\omega') 
\nonumber \\ & \quad \times
  \frac{\muz\hbar\omega^2}{\ii2\pi}
  \left[ \mGren(\vr,\vr',\omega) - \cjgt{\mGren(\vr',\vr,\omega)} \right],
\label{eq:[hvE+,hvE-]-mGren} \\ 
& \left[ \hvE^+(\vr,\omega), \hvE^+(\vr',\omega') \right] = \mzero,
\end{align}
\end{subequations}
where $\mGren(\vr,\vr',\omega)$ is defined as
\begin{align}&
\mGren(\vr,\vr',\omega) = \mG(\vr,\vr',\omega)
\nonumber \\ & \quad
+ \frac{1}{\muz\omega^2} \sum_{\mu,\mu'} \vdimE_{\mu}(\vr,\omega)\
  \crd_{\mu,\mu'}(\omega)\ \vdimF_{\mu'}(\vr',\omega). \label{eq:rep-mGren} 
\end{align}
We can find that if this tensor satisfies the equation
\begin{align}&
\rot\rot\mGren(\vr,\vr',\omega)
- \frac{\omega^2}{c^2}\dieb(\vr,\omega)\mGren(\vr,\vr',\omega)
\nonumber \\ &
- \frac{\omega^2}{c^2} \int\dd\vr''\ \mchiex(\vr,\vr'',\omega)
  \cdot \mGren(\vr'',\vr',\omega)
= \delta(\vr-\vr')\munit,
\label{eq:satisfied-mGren} 
\end{align}
then it can be interpreted as Green's tensor for nonlocal Maxwell wave 
equation \eqref{eq:Maxwell-nonlocal}.
This tensor, which renormalizes the linear optical process of excitons 
with the microscopic nonlocality, is also calculated directly from the above
nonlocal equation.\cite{cho03jl}
The breakdown of reciprocity relation \eqref{eq:recip-mG}
for $\mGren(\vr,\vr',\omega)$
arises from anisotropic susceptibility tensor \eqref{eq:def-chi-ex}
of the excitonic polarization.
By using $\mGren(\vr,\vr',\omega)$, 
electric field operator \eqref{eq:hvEone=hvEz+E*hexone} is also
written as
\begin{equation} \label{eq:rep-hvE-mGren-hvJz} 
\hvE^+(\vr,\omega)
= \ii\muz\omega \int\dd\vr'\ \mGren(\vr,\vr',\omega) \cdot \hvJz(\vr',\omega).
\end{equation}
Comparing Eqs.~\eqref{eq:E=GJ-bg}, \eqref{eq:[E(r,w),E(r,w)]-bg},
\eqref{eq:[hvEone,hvEone]}, and \eqref{eq:rep-hvE-mGren-hvJz},
we can find a good correspondence between our theory and the QED theories
for local systems.
In addition, we also verify that commutation relations 
\eqref{eq:[hexone,hexone]} and \eqref{eq:[hvEone,hvEone]} satisfy 
the equal-time commutation relations expected 
in the Shr\"{o}dinger representation (see App.~\ref{app:stcr}).

\section{\label{sec:absorption}With nonradiative relaxation}
Next, we discuss the modulation of the above thoery
by considering the nonradiative relaxation of excitons.
The detailed caluclation is shown in App.~\ref{sec:with-abs},
and the result is as follows.
The nonradiative relaxation slightly modulates
the motion equation of excitons from the original one \eqref{eq:motion-hexone}:
\begin{align} &
\left[ \hbar\wex_{\mu}-\hbar\omega-\ii\damp_{\mu}(\omega)/2 \right] 
\hex_{\mu}(\omega)
\nonumber \\ &
= \int\dd\vr\ \vdimP_{\mu}^*(\vr) \cdot \hvE^+(\vr,\omega)
+ \hrsrc_{\mu}(\omega).
\label{eq:motion-hexone-damp} 
\end{align}
$\damp_{\mu}(\omega)$ is the nonradiative relaxation width defined 
as Eq.~\eqref{eq:def-gamma} in terms of exciton-reservoir interaction
coefficient $\ccxr_{\mu}(\wres)$.
Operator $\hrsrc_{\mu}(\omega)$ represents the fluctuation caused
by the reservoir oscillators. 
It is defined as Eq.~\eqref{eq:def-hrsrc} 
and satisfies the commutation relations
\begin{subequations} \label{eq:[hrsrc,hrsrc]} 
\begin{align}
\left[ \hrsrc_{\mu}(\omega), \dgg{\hrsrc_{\mu'}(\omega')} \right]
& = \delta_{\mu,\mu'} \delta(\omega-\omega') \frac{\hbar}{2\pi}
  \damp_{\mu}(\omega), 
\label{eq:[hrsrc,hrsrcd]} \\ 
\left[ \hrsrc_{\mu}(\omega), \hrsrc_{\mu'}(\omega') \right] & = 0.
\label{eq:[hrsrcd,hrsrcd]} 
\end{align}
\end{subequations}
This is another source operator of our system
and is independent from noise current density $\hvJz(\vr,\omega)$ as
\begin{equation} \label{eq:[hrsrc,hvJz]} 
\left[ \hrsrc_{\mu}(\omega), \hvJz(\vr,\omega') \right]
= \left[ \hrsrc_{\mu}(\omega), \dgg{\hvJz(\vr,\omega')} \right]
= \vzero.
\end{equation}
Substituting representation \eqref{eq:E=Ez+G*Pex} of the electric field
into Eq.~\eqref{eq:motion-hexone-damp}, we obtain the self-consistent equation
set as
\begin{equation} \label{eq:self-consist-damp} 
\sum_{\mu'} \cssa_{\mu,\mu'}(\omega) \hex_{\mu'}(\omega)
= \int\dd\vr\ \vdimP_{\mu}^*(\vr) \cdot \hvEz^+(\vr,\omega)
+ \hrsrc_{\mu}(\omega),
\end{equation}
where the coefficient matrix element is
\begin{equation} \label{eq:def-S-damp} 
\cssa_{\mu,\mu'}(\omega)
\equiv \left[ \hbar\wex_{\mu}-\hbar\omega-\ii\damp_{\mu}(\omega)/2 \right]
 \delta_{\mu,\mu'} + \rct_{\mu,\mu'}(\omega).
\end{equation}

The above result indicates that we can easily introduce 
the nonradiative relaxation into the discussion of the previous section.
All we have to do is introduce relaxation width $\damp_{\mu}(\omega)$
into coefficient matrix \eqref{eq:def-S-damp}
of the self-consistent equation set.
By deriving its inverse matrix $\mcrda(\omega) = [\mcssa(\omega)]^{-1}$, 
we obtain the commutation relations for excitons as
\begin{subequations} \label{eq:[hexone,hexone]-damp} 
\begin{align}&
\left[ \hex_{\mu}(\omega), \dgg{\hex_{\mu'}(\omega')} \right]
\nonumber \\ & \quad
= \delta(\omega-\omega')\ \frac{\hbar}{\ii2\pi}
  \left[ \crda_{\mu,\mu'}(\omega) - \cjg{\crda_{\mu',\mu}(\omega)} \right], \\
& \left[ \hex_{\mu}(\omega), \hex_{\mu'}(\omega') \right]
= 0.
\end{align}
\end{subequations}
These have the same forms as those of Eqs.~\eqref{eq:[hexone,hexone]}.

On the other hand, substituting Eq.~\eqref{eq:motion-hexone-damp} 
into Eq.~\eqref{eq:Pex=P*ex-rwa}, 
the excitonic polarization \eqref{eq:Pexone=chi*E} is rewritten as
\begin{align} \label{eq:rep-hvPex-damp} 
\hvPex^+(\vr,\omega) 
& = \diez \int\dd\vr'\ \mchiexa(\vr,\vr',\omega) \cdot \hvE^+(\vr,\omega)
\nonumber \\ & \quad
  + \sum_{\mu} \frac{\vdimP_{\mu}(\vr)\hrsrc_{\mu}(\omega)}
                    {\hbar\omega_{\mu}-\hbar\omega-\ii\damp_{\mu}(\omega)/2},
\end{align}
where the nonlocal susceptibility is also rewritten as
\begin{equation} \label{eq:def-mchiexa} 
\mchiexa(\vr,\vr',\omega) \equiv \frac{1}{\diez} \sum_{\mu}
  \frac{\vdimP_{\mu}(\vr)\vdimP_{\mu}^*(\vr')}
       {\hbar\wex_{\mu}-\hbar\omega-\ii\damp_{\mu}(\omega)/2}.
\end{equation}
Substituting Eq.~\eqref{eq:rep-hvPex-damp} into Maxwell wave equation
\eqref{eq:Maxwell-E-Jz-Pex}, nonlocal wave equation 
\eqref{eq:Maxwell-nonlocal} becomes
\begin{align}&
\rot\rot\hvE^+(\vr,\omega)
- \frac{\omega^2}{c^2}\dieb(\vr,\omega)\hvE^+(\vr,\omega)
\nonumber \\ &
- \frac{\omega^2}{c^2} \int\dd\vr'\ \mchiexa(\vr,\vr',\omega)
  \cdot \hvE^+(\vr',\omega)
= \ii\muz\omega\hvJza(\vr,\omega),
\label{eq:Maxwell-nonlocal-damp2} 
\end{align}
where we define a new noise current density as
\begin{equation}
\hvJza(\vr,\omega) \equiv \hvJz(\vr,\omega) 
- \ii\omega \sum_{\mu}
  \frac{\vdimP_{\mu}(\vr)\hrsrc_{\mu}(\omega)}
       {\hbar\omega_{\mu}-\hbar\omega-\ii\damp_{\mu}(\omega)/2}.
\end{equation}
From commutation relations \eqref{eq:[hvJz,hvJz]}, 
\eqref{eq:[hrsrc,hrsrc]}, and \eqref{eq:[hrsrc,hvJz]},
we can obtain the commutation relations for $\hvJza(\vr,\omega)$ as
\begin{subequations} \label{eq:[hvJza,hvJza]} 
\begin{align}
&   \left[ \hvJza(\vr,\omega), \dgg{\hvJza(\vr',\omega')} \right] \nonumber \\
& \quad = \delta(\omega-\omega') \frac{\diez\hbar\omega^2}{\pi}
    \Im[\mdiea(\vr,\vr',\omega)] \\
& \left[ \hvJza(\vr,\omega), \hvJza(\vr',\omega') \right] = \mzero,
\end{align}
\end{subequations}
where $\mdiea(\vr,\vr',\omega) 
\equiv \munit\delta(\vr-\vr')\dieb(\vr,\omega) + \mchiexa(\vr,\vr',\omega)$
is the nonlocal dielectric tensor.
This is a natural result from the fluctuation theorem
as discussed in Ref.~\onlinecite{stefano01}-\onlinecite{raabe07}.
In addition, we also obtain the commutation relatons for the electric field as 
having the same form as that of Eq.~\eqref{eq:[hvEone,hvEone]}
by replacing $\mGren(\vr,\vr',\omega)$ with
\begin{align}&
\mGrena(\vr,\vr',\omega) = \mG(\vr,\vr',\omega)
\nonumber \\ & \quad
+ \frac{1}{\muz\omega^2} \sum_{\mu,\mu'} \vdimE_{\mu}(\vr,\omega)\
  \crda_{\mu,\mu'}(\omega)\ \vdimF_{\mu'}(\vr',\omega),
\label{eq:rep-mGren-damp} 
\end{align}
which satisfies the nonlocal wave equation
\begin{align}&
\rot\rot\mGrena(\vr,\vr',\omega)
- \frac{\omega^2}{c^2}\dieb(\vr,\omega)\mGrena(\vr,\vr',\omega)
\nonumber \\ &
- \frac{\omega^2}{c^2} \int\dd\vr''\ \mchiexa(\vr,\vr'',\omega)
  \cdot \mGrena(\vr'',\vr',\omega)
= \delta(\vr-\vr')\munit
\label{eq:Maxwell-nonlocal-damp} 
\end{align}
and Green's tensor required 
for the practical application of nonlocal QED theories.
\cite{savasta02pra,raabe07}

\section{\label{sec:discussion}Discussion}
In this paper, we have constructed a QED theory for excitons
weakly confined in arbitrary-structured dielectrics
with microscopic nonlocality and nonradiative relaxation of excitons.
On the other hand, as mentioned in Sec.~\ref{sec:nonlocalQED}, 
the QED of media with the microscopic nonlocality
has already been discussed in a few studies.
From the viewpoint of a practical application,
we compare our theory with the studies
of Stefano et al.\cite{savasta02pra,stefano01,stefano99}
and Raabe et al.\cite{raabe07}

Stefano et al.~have discussed
quantum-well structures of the dispersive and absorptive dielectrics 
with microscopic nonlocality in Ref.~\onlinecite{stefano99},
and their theory is generalized to enable the consideration of
arbitrary structures
in Ref.~\onlinecite{stefano01} and \onlinecite{savasta02pra}.
However, there still remains a problem in deriving Green's tensors for
the nonlocal wave equation as shown in Eq.~\eqref{eq:satisfied-mGren}
or \eqref{eq:Maxwell-nonlocal-damp} in the present paper.
Our thoery provides a solution to this problem
by giving a definite calculation method
using Green's tensor satisfying wave equation 
\eqref{eq:satisfied-mG} for local media and 
the fact that the nonlocal susceptibility has a separable form
as shown in Eq.~\eqref{eq:def-chi-ex} or \eqref{eq:def-mchiexa}.
The problem of Ref.~\onlinecite{savasta02pra} can be solved 
by using our theory
because we derive Green's tensor \eqref{eq:rep-mGren}
or \eqref{eq:rep-mGren-damp}
for arbitrary-structured excitonic polarization and background dielectrics.

On the other hand, Raabe et al.~propose the use of the dielectric approximation
with the surface impedance method for the practical calculation
of Green's tensor for the nonlocal Maxwell wave equation.
In the dielectric apprximation,
the characteristic length of spatial dispersion
(the spatial spreading of the center-of-mass motion of excitons)
is assumed to be small compared to the spatial length of materials,
and the information outside a focusing region is compressed to 
integrations of the electromagnetic fields at the interfaces.
Green's tensor can be derived using the surface impedance method
for a given surface impedance or admittance,
which just includes the outside information.
In contrast, our thoery provides Green's tensor
for given $\dieb(\vr,\omega)$ and center-of-mass wave functions of excitons
by applying only the RWA.
Since nonlocality becomes essentail only under the resonance conditions,
the RWA does not impose any significant restriction on our thoery 
for the discussion of nonlocal systems.
In addition, our calculaton method would be generalized 
to the one without the RWA
as performed in the semiclassical nonlocal theory.\cite{cho91,cho03}

As mentioned in Sec.~\ref{sec:intro},
there is grwoing interest in the QED of elementary excitations 
in condensed matters.
For example, theoretical studies on entangled-photon generation 
via biexcitons have already been performed
by Savasta et al.\cite{savasta99prb,savasta99ssc}
(though the microscopic nonlocality was not sufficiently considered 
in these calculations)
and also by us by extending our QED theory reported in the present paper.
\cite{bamba06}
When we discuss such nonlinear processes of excitons,
we must self-consistently treat
their nonlinear motion equation and the Maxwell wave equation.
Based on self-consistent equation set \eqref{eq:self-consist-S}
or \eqref{eq:self-consist-damp} as discussed in the present paper,
the new task is to solve the equation set with nonlinear terms
originating from nonlinear processes.
On the other hand, based on Maxwell wave equation 
\eqref{eq:Maxwell-nonlocal} or \eqref{eq:Maxwell-nonlocal-damp2} 
with the nonlocal susceptibility as discussed
in the previously discussed QED theories,
\cite{stefano99,stefano01,savasta02pra,bechler06,suttorp07,raabe07}
we must solve the wave equation with both nonlinear 
and nonlocal susceptibility.
Both approaches can be performed by applying some techniques
such as successive approximation;
however, they generally require much hard work.
In such a case, more detailed and systematic calculation should be performed
by using the Feynman diagram technique 
with the correlation functions of excitons derived using our QED theory.
In this sense, our scheme will be a powerfull tool to discuss
the nonlinear quantum optics in condensed matters
with the microscopic nonlocality.

\section{\label{sec:summary}Summary}
By using the quantization technique of Suttorp et al.,\cite{wubs01,suttorp04} 
we have extended the microscopic nonlocal theory\cite{cho91,cho03} developed
in the semiclassical framework to enable the consideration of
quantum mechanical properties of the electromagnetic fields.
This theory keeps good correspondences with both the nonlocal theory 
and the QED theories
for dispersive and absorptive materials with local susceptibility.
While microscopic nonlocality has been considered in the form of 
nonlocal susceptibility in some of the QED theories,
\cite{stefano99,stefano01,savasta02pra,suttorp07,raabe07}
we reduce the problem to a linear equation set
as discussed in the semiclassical framework.\cite{cho91,cho03}
In addition, this theory provides Green's tensor for the Maxwell wave
equation with a nonlocal susceptibility,
which is just required for a practical application 
of previous studies.
By using our theory, we can discuss the QED of excitons weakly confined
in nano-structures, which are known to show anomalous nonlinear optical
phenomena.\cite{ishihara96,akiyama99,ishihra02jan,ishihara02jul,ishihara03}
Although this paper is devoted to only the linear optical process of excitons,
we can phenomenologically extend our theory to describe nonlinear processes,
such as the entangled-photon generation via biexcitons 
in nano-structures.\cite{bamba06}
In addition, our theory has a potential to systematically discuss higher order
nonlinear processes of elementary excitations in condensed matters
by using the Feynman diagram technique with correlation functions of excitons
derived in our calculation.
Based on our QED theory, we are going to discuss various optical phenomena
which cannot be discussed in the semiclassical framework.

\begin{acknowledgments}
The authors are grateful to Prof.~K.~Cho, Dr.~H.~Ajiki, and Dr.~K.~Koshino 
for helpful discussions.
This work was partially supported by
the Japan Society for the Promotion of Science (JSPS);
a Grant-in-Aid for Creative Science Research, 17GS1204, 2005;
and JSPS Research Fellowships for Young Scientists.
\end{acknowledgments}

\appendix
{
\def\ovJcp{\mathbf{J}_{\text{cp}}}
\def\ovIcp{\mathbf{I}_{\text{cp}}}
\def\ovPcp{\mathbf{P}_{\text{cp}}}
\def\oNcp{N_{\text{cp}}}
\def\orhocp{\rho_{\text{cp}}}
\def\ophicp{\phi_{\text{cp}}}

\def\ovPex{\hat{\mathbf{P}}_{\text{ex}}}
\def\ovJex{\hat{\mathbf{J}}_{\text{ex}}}
\def\ovIex{\hat{\mathbf{I}}_{\text{ex}}}
\def\oNex{\hat{N}_{\text{ex}}}
\def\orhoex{\hat{\rho}_{\text{ex}}}
\def\ophiex{\hat{\phi}_{\text{ex}}}

\def\uvol{\Omega}
\def\intu{\int_{\Omega}}
\def\blf{\varphi}
\def\wnf{w}
\def\wvfpcl{\varphi}

\def\oH{\hat{H}}
\def\oHmat{\hat{H}_{\mathrm{mat}}}
\def\ofld{\hat{\psi}}
\def\ofldd{\hat{\psi}^{\dagger}}
\def\opcl{\hat{a}}
\def\opcld{\hat{a}^{\dagger}}
\def\oex{\hat{b}}

\section{The second quantization of excitonic polarization\label{app:2ndJP}}
In this appendix, we provide microscopic descriptions 
of the current density, charge density, and polarization density
of charged particles.
Then, we expand them in terms of the electron or exciton operators.
We write the second-quantized operators with a hat ( $\hat{}$ ) 
in this appendix.

Considering the charged particles with mass $m_i$ and charge $q_i$
at position $\vr_i$,
current density $\ovJcp(\vr)$ and 
charge density $\orhocp(\vr)$ are written as
\begin{align}
\ovJcp(\vr) & \equiv \sum_i \frac{q_i}{2}
\left[ \dot{\vr}_i \delta(\vr-\vr_i) + \delta(\vr-\vr_i) \dot{\vr}_i \right],
\label{eq:def-cur-micro} \\ 
\orhocp(\vr) & \equiv \sum_i q_i \delta(\vr-\vr_i).
\label{eq:def-dens-micro} 
\end{align}
Here, due to the interaction with the radiation field
(see App.~\ref{app:hamilt}), 
the momentum of the charged particles is written as
\begin{equation}
\vp_i = m_i \dot{\vr}_i + q_i \ovA(\vr_i).
\end{equation}
Then, current density \eqref{eq:def-cur-micro} includes 
a contribution from the radiation field.
In order to expand it in terms of electron or exciton operators,
we extract the radiation contribution from $\ovJcp(\vr)$:
\begin{equation} \label{eq:def-I-micro} 
\ovIcp(\vr) \equiv \sum_i \frac{q_i}{2m_i}
\left[ \vp_i \delta(\vr-\vr_i) + \delta(\vr-\vr_i) \vp_i \right].
\end{equation}
By writing the coefficient of the radiation contribution as
\begin{equation} \label{eq:def-N-micro} 
\oNcp(\vr) \equiv \sum_i \frac{{q_i}^2}{m_i} \delta(\vr-\vr_i),
\end{equation}
we can write complete current density \eqref{eq:def-cur-micro} as
\begin{align}
\ovJcp(\vr)
& = \ovIcp(\vr) - \oNcp(\vr) \ovA(\vr).
\end{align}
This subtraction of the radiation contribution
is discussed in section 2.2 of Ref.~\onlinecite{cho03}.

Next, we expand the above variables
in terms of the electron operator $\opcl_{\eta}$ 
and its wave function $\wvfpcl_{\eta}(\vr)$.
The field operator is written as
\begin{equation} \label{eq:def-ofld} 
\ofld(\vr) = \sum_{c} \opcl_c \wvfpcl_c(\vr) + \sum_v \opcl_v \wvfpcl_v(\vr),
\end{equation}
where labels $c$ and $v$ represent the degrees of freedom of 
conduction and valence electrons, respectively.
Assuming optical excitation of electron-hole pairs,
we obtain the second-quantized form of the above variables as
\begin{align}
\ovIex(\vr)
& = \frac{(-e)}{2m} \sum_{c,v} \opcld_{v} \opcl_{c}
    \left[ \wvfpcl_{v}^*(\vr) \vp \wvfpcl_{c}(\vr) 
         - \wvfpcl_{c}(\vr) \vp \wvfpcl_{v}^*(\vr) \right]
\nonumber \\ & \quad
  + \Hc, \\
\oNex(\vr)
& = \frac{(-e)^2}{m} \sum_{c,v} \opcld_{v} \opcl_{c} 
    \wvfpcl_{v}^*(\vr) \wvfpcl_{c}(\vr) + \Hc, \\
\orhoex(\vr) 
& = (-e) \sum_{c,v} \opcld_{v} \opcl_{c} 
    \wvfpcl_{v}^*(\vr) \wvfpcl_{c}(\vr) + \Hc
\label{eq:rep-orhoex-opcl} 
\end{align}
These operators are also written in terms of exciton operators
$\{\oex_{\mu}\}$ as polarization density \eqref{eq:rep-ovPex-oex}:
\begin{align}
\ovIex(\vr) & = \sum_{\mu}\vdimI_{\mu}(\vr) \oex_{\mu}+ \Hc, 
\label{eq:rep-ovIex-ex} \\ 
\oNex(\vr) & = \sum_{\mu}N_{\mu}(\vr) \oex_{\mu}+ \Hc,
\label{eq:rep-oNex-ex} \\ 
\orhoex(\vr) & = \sum_{\mu}\rho_{\mu}(\vr) \oex_{\mu}+ \Hc
\label{eq:rep-orhoex-ex} 
\end{align}
Instead of evaluating the expansion coefficient of each operator,
we describe them in terms of $\vdimP_{\mu}(\vr)$, 
the coefficient of polarization density \eqref{eq:rep-ovPex-oex}.
From the relations
\begin{align}
\orhoex(\vr) & = - \div \ovPex(\vr),
\label{eq:orhoex=div*ovPex} \\ 
\ovJex(\vr)
& = \ddt{}\ovPex(\vr)
  = \frac{1}{\ii\hbar} \left[ \ovPex(\vr), \oH \right],
\label{eq:ovJex=ddt*ovPex} 
\end{align}
and considering weak exciton-photon interaction,
i.e., $\oH \sim \oHmat$ and $\ovJex(\vr) \sim \ovIex(\vr)$, 
we can write the above coefficients as
\begin{align}
\vdimI_{\mu}(\vr) & = - \ii \wex_{\mu}\vdimP_{\mu}(\vr),
\label{eq:vdimI=w*vdimP} \\ 
N_{\mu}(\vr) & = (-e/m) \rho_{\mu}(\vr), \\
\rho_{\mu}(\vr) & = - \div\vdimP_{\mu}(\vr), \label{eq:rho=div*vdimP} 
\end{align}
where $\wex_{\mu}$ is the eigenfrequency of excitons.
Using above operators \eqref{eq:rep-ovIex-ex}-\eqref{eq:rep-orhoex-ex},
the excitonic current density and Coulomb potential are respectively
written as
\begin{align}
\ovJex(\vr) & = \ovIex(\vr) - \oNex(\vr)\ovA(\vr), 
\label{eq:rep-ovJex} \\ 
\ophiex(\vr) & = \int\dd\vr'\ \frac{\orhoex(\vr')}{4\pi\diez|\vr-\vr'|}.
\label{eq:rep-ophiex} 
\end{align}

In order to evaluate coefficients 
\eqref{eq:vdimI=w*vdimP}-\eqref{eq:rho=div*vdimP},
we derive the representation of $\vdimP_{\mu}(\vr)$
from the microscopic description of the polarization density.
Averaging the polarization at lattice point $\vRz$ over a unit cell,
the microscopic description is written as
\begin{equation}
\ovPcp(\vRz) \equiv \frac{1}{\uvol} \intu\dd\vr 
\sum_i q_i\vr\ \delta(\vRz+\vr-\vr_i),
\label{eq:def-plz-micro} 
\end{equation}
where the integration is over the unit cell and $\uvol$ is its volume.
Explicitly indicating the lattice point of the electron states
as $(\eta, \vR)$ and assuming their wave function 
as the wannier function $\wnf_{\eta}(\vr-\vR)$,
we obtain the second-quantized form of the polarization density as
\begin{align} &
\ovPex(\vRz)
= \sum_{c,v,\vR,\vR'} \opcld_{v,\vRz+\vR'} \opcl_{c,\vRz+\vR+\vR'}
\nonumber \\ & \quad \times
  \frac{1}{\uvol}\intu\dd\vr\
  \wnf_{v}^*(\vr-\vR')\ (-e)\vr\ \wnf_{c}(\vr-\vR-\vR')
+ \Hc
\label{eq:rep-ovPex-eh} 
\end{align}
We expand this in terms of the exciton operators
\begin{equation} \label{eq:def-ex-opr} 
\oex_{\mu,m}
\equiv \sum_{c,v,\vRz,\vR} \rmfex_{\mu,c,v,\vR}^*\ \cmfex_{m,\vRz}^*\
  \opcld_{v,\vRz}\ \opcl_{c,\vRz+\vR},
\end{equation}
where $\mu$ and $\rmfex_{\mu,c,v,\vR}$ are, respectively, the quantum number 
and the wave function of the relative motion of excitons,
and $m$ and $\cmfex_{m,\vRz}$ are those of the center-of-mass motion.
From the completeness of the wave functions, we can rewrite the electron-hole 
operator set as
\begin{equation} \label{eq:eh-to-ex} 
\opcld_{v,\vRz}\ \opcl_{c,\vRz+\vR}
= \sum_{\mu,m} \rmfex_{\mu,c,v,\vR}\ \cmfex_{m,\vRz}\ \oex_{\mu,m}.
\end{equation}
Using this relation, we can expand Eq.~\eqref{eq:rep-ovPex-eh}
in terms of the exciton operators as
\begin{equation} \label{eq:rep-ovPex-ex(mu,m)} 
\ovPex(\vRz) = \sum_{\mu,m} \vdimP_{\mu,m}(\vRz) \oex_{\mu,m} + \Hc,
\end{equation}
where the expansion coefficient is written as
\begin{align} &
\vdimP_{\mu,m}(\vRz)
\equiv \sum_{c,v,\vR,\vR'} \cmfex_{m,\vRz+\vR'}\ \rmfex_{\mu,c,v,\vR}
\nonumber \\ & \times
  \frac{1}{\uvol}\intu\dd\vr\
  \wnf_{v}^*(\vr-\vR')\ (-e)\vr\ \wnf_{c}(\vr-\vR-\vR').
\label{eq:coeff-P-Wannier}
\end{align}
Supposing that the spatial variation of the center-of-mass wave function
is negligible within the extent of the electron-hole relative wave function,
we can consider 
$\cmfex_{m,\vRz+\vR'} \sim \cmfex_{m,\vRz}$.
Further, by expanding the integration range to the entire crystal region
with iterating $\vR'$, we obtain
\begin{align}
\vdimP_{\mu,m}(\vRz)
& \sim \cmfex_{m,\vRz}\sum_{c,v,\vR}\rmfex_{\mu,c,v,\vR}
\nonumber \\ & \quad \times
  \frac{1}{\uvol}\int\dd\vr\
  \wnf_{v}^*(\vr)\ (-e)\vr\ \wnf_{c}(\vr-\vR).
\end{align}
Here, if we assume the wave functions to be smooth 
with respect to the spatial position, i.e., 
$\cmfex_{m}(\vRz) = \cmfex_{m,\vRz} / \sqrt{\uvol}$ and
$\rmfex_{\mu,c,v}(\vR) = \rmfex_{\mu,c,v,\vR} / \sqrt{\uvol}$,
we obtain the expansion coefficient of polarization density 
\eqref{eq:rep-ovPex-ex(mu,m)} as
\begin{equation} 
\vdimP_{\mu,m}(\vRz) = \vdimP_{\mu}\ \cmfex_{m}(\vRz),
\end{equation}
where
\begin{equation}\label{eq:coeff-P-rm}
\vdimP_{\mu}
\equiv \sum_{c,v,\vR}\rmfex_{\mu,c,v}(\vR)\
  \int\dd\vr\ \wnf_{v}^*(\vr)\ (-e)\vr\ \wnf_{c}(\vr-\vR)
\end{equation}
is the transition dipole moment of exciton band $\mu$,
and its absolute value is related with the LT splitting 
of the exciton eigenenergy as $\DLT^{\mu} = |\vdimP_{\mu}|^2/\diez\dieb$.
}
{
\def\longi#1{\left[#1\right]_{\text{L}}}
\def\trans#1{\left[#1\right]_{\text{T}}}
\def\wz{\omega_0}
\def\wzt{\tilde{\omega}_0}

\def\vA{\mathbf{A}}
\def\vB{\mathbf{B}}

\def\ofld{\psi}
\def\ofldd{\psi^{\dagger}}
\def\opcl{a}
\def\opcld{a^{\dagger}}
\def\wvfpcl{\varphi}

\def\Lag{L}
\def\dLag{\mathcal{L}}
\def\vPi{\mathbf{\Pi}}
\def\vX{\mathbf{X}}
\def\vP{\mathbf{P}}
\def\vQ{\mathbf{Q}}
\def\vY{\mathbf{Y}}
\def\rhobg{\rho_{\text{bg}}}
\def\phibg{\phi_{\text{bg}}}
\def\rhocp{\rho_{\text{cp}}}
\def\phicp{\phi_{\text{cp}}}
\def\vJcp{\mathbf{J}_{\text{cp}}}

\def\ovJexT{\mathbf{J}_{\text{exT}}}
\def\ovJexL{\mathbf{J}_{\text{exL}}}
\def\ovPi{\mathbf{\Pi}}
\def\ovX{\mathbf{X}}
\def\ovP{\mathbf{P}}
\def\ovQ{\mathbf{Q}}
\def\ovY{\mathbf{Y}}
\def\oOmega{\Omega}

\def\ovJcp{\mathbf{J}_{\text{cp}}}
\def\ovIcp{\mathbf{I}_{\text{cp}}}
\def\ovPcp{\mathbf{P}_{\text{cp}}}
\def\oNcp{N_{\text{cp}}}
\def\orhocp{\rho_{\text{cp}}}
\def\ophicp{\phi_{\text{cp}}}

\def\dovJex{\dot{\mathbf{J}}_{\text{ex}}}
\def\dovPexL{\dot{\mathbf{P}}_{\text{exL}}}
\def\dorhoex{\dot{\rho}_{\text{ex}}}
\def\dophiex{\dot{\phi}_{\text{ex}}}

\section{Derivation of Hamiltonian\label{app:hamilt}}
As a model of the background system, 
i.e., radiation field and local dielectrics, 
we adopt the system discussed by SW.\cite{suttorp04}
Considering the charged particles of App.~\ref{app:2ndJP},
the total Lagrangian is written as
\begin{equation} \label{eq:Lagrangian} 
\Lag = \sum_i \left[ \half m_i {\dot{\vr}_i}^2 - V(\vr_i) \right]
     + \int\dd\vr\ \dLag,
\end{equation}
where $V(\vr_i)$ is the one-body potential of the particles
and $\dLag$ is the Lagrangian density depending on the spatial position:
\begin{align} \label{eq:dLag} 
\dLag 
& = \half \diez \vE^2 - \frac{1}{2\muz} \vB^2
  + \half \rho \dot{\vX}^2 - \half \rho{\wz}^2 \vX^2
\nonumber \\ & \quad 
  - ( \phibg + \phicp ) ( \rhobg + \rhocp )
  + \vA \left( -\alpha\dot{\vX} + \vJcp \right)
\nonumber \\ & \quad 
  + \half \rho \int_0^{\infty}\dd\omega\ {\dot{\vY}_{\omega}}^2
  - \half \rho \int_0^{\infty}\dd\omega\ \omega^2 {\vY_{\omega}}^2
\nonumber \\ & \quad 
  - \int_0^{\infty}\dd\omega\ v_{\omega} \vX \cdot \dot{\vY}_{\omega}.
\end{align}
We omit the descriptions of position dependences.
$\vE = - \dot{\vA} - \grad\phibg - \grad\phicp$ is the electric field,
and $\vB = \rot \vA$ is the magnetic induction.
$\ovX(\vr)$ is the amplitude of polarizable harmonic oscillators
with density $\rho(\vr)$ and eigenfrequency $\wz(\vr)$.
These oscillators describe the background medium in our QED theory.
The polarization density, charge density, 
and current density of the background are respectively written as
$-\alpha\ovX$, $\orhobg = \div(\alpha\ovX)$, and $-\alpha\dot{\ovX}$
with position-dependent coefficient $\alpha(\vr)$.
The background Coulomb potential is written as
\begin{equation}
\ophibg = \int\dd\vr'\ \frac{\orhobg'}{4\pi\diez|\vr-\vr'|}
= \int\dd\vr'\ \frac{\vnabla'\cdot(\alpha'\ovX')}{4\pi\diez|\vr-\vr'|}.
\end{equation}
$\ophibg$ is related with the longitudinal component
of the polarization as
\begin{equation} \label{eq:plz-L-bg} 
\grad\ophibg = - \frac{1}{\diez}\longi{\alpha\ovX},
\end{equation}
and it also satisfies the Poisson equation
\begin{equation} \label{eq:Poisson} 
\Lapl\ophibg = - \frac{\orhobg}{\diez}
= - \frac{1}{\diez}\div(\alpha\ovX).
\end{equation}
The damping in the background system is described by a reservoir of oscillators
interacting with the polarizable ones.
$\ovY_{\omega}(\vr)$ is the amplitude of the oscillators
with frequency $\omega$, 
and $v_{\omega}(\vr)$ represents the coupling strength.

From Lagrangian \eqref{eq:Lagrangian}, the canonical momenta of the above
variables are derived as
\begin{subequations}
\begin{align}
\vPi & \equiv \frac{\partial \Lag}{\partial \dot{\vA}} 
       = \diez \dot{\vA}, \\
\vP & \equiv \frac{\partial \Lag}{\partial \dot{\vX}} 
      = \rho \dot{\vX} - \alpha\vA, \label{eq:P=V-A} \\ 
\vQ_{\omega} & \equiv \frac{\partial \Lag}{\partial \dot{\vY}_{\omega}} 
               = \rho \dot{\vY}_{\omega} - v_{\omega} \vX, \\
\vp_i & \equiv \frac{\partial \Lag}{\partial \dot{\vr}_i}
        = m_i \vr_i + q_i \vA(\vr_i). \label{eq:p=mv+A} 
\end{align}
\end{subequations}
Since $\vA$ and $\grad(\phibg\ +\ \phicp)$ are perpenticular to each other,
from the Poisson equation \eqref{eq:Poisson} and $\Lapl\phicp = -\rhocp/\diez$,
the first term of Eq.~\eqref{eq:dLag} is rewritten as
\begin{equation}
\int\dd\vr\ \frac{\diez}{2} \vE^2
= \int\dd\vr \left[ \frac{\vPi^2 }{2\diez}
+ \half (\phibg+\phicp)(\rhobg+\rhobg) \right].
\end{equation}
After a straightforward calculation, we obtain the Hamiltonian as
\begin{align} \label{eq:H-cp} 
\oH & = \oHem
+ \sum_i \left[ \frac{1}{2m_i} \left\{ \vp_i-q_i\vA(\vr_i) \right\}^2 
            + V(\vr_i) \right]
\nonumber \\ & \quad
+ \int\dd\vr\ \left[ \half \phicp \rhocp + \phibg \rhocp \right],
\end{align}
where $\oHem$ describes the complete Hamiltonian discussed by SW,
representing 
the radiation field and background dielectrics with local susceptibility:
\begin{align} \label{eq:def-Hem} 
\oHem
& = \int\dd\vr \biggl[ 
    \frac{\ovPi^2}{2\diez}
  + \frac{1}{2\muz}(\rot\ovA)^2
  + \frac{\ovP^2}{2\rho}
  + \frac{\rho\wzt{}^2}{2} \ovX^2
\nonumber \\ & \quad
  + \int_0^{\infty}\dd\omega\ \frac{{\ovQ_{\omega}}^2}{2\rho}
  + \int_0^{\infty}\dd\omega\ \frac{\rho\omega^2}{2}{\ovY_{\omega}}^2
  + \frac{\alpha}{\rho}\ovP\cdot\ovA
\nonumber \\ & \quad
  + \frac{\alpha^2}{2\rho}\ovA^2
  + \int_0^{\infty}\dd\omega\ \frac{v_{\omega}}{\rho}\ovX\cdot\ovQ_{\omega}
  + \half \phibg \rhobg
  \biggr].
\end{align}
The first two terms represent the radiation energy,
the third term is the kinetic energy of the oscillators,
and the fourth is the potential.
The seventh and eighth terms represent the interaction between the oscillators
and the radiation field.
The eigenfrequency of the oscillators shown
in the forth term of Eq.~\eqref{eq:def-Hem}
is modified as
\begin{equation}
\wzt{}^2
\equiv {\wz}^2 + \frac{1}{\rho}\int_0^{\infty}\dd\omega\ {v_{\omega}}^2
\end{equation}
by the interaction with the reservoir oscillators,
which is described as the ninth term.
The energy of reservoir is the fifth and sixth terms.
The last term is the Coulomb interaction between the induced charges 
of backgrounds.

The kinetic energy of the charged particles, 
the second term of Eq.~\eqref{eq:H-cp}, is expanded
with the expression \eqref{eq:p=mv+A} of their momentum as
\begin{align}&
\sum_i \frac{1}{2m_i}\left[ \vp_i - q_i\ovA(\vr_i) \right]^2
\nonumber \\ &
= \sum_i \frac{1}{2m_i}{\vp_i}^2
- \sum_i \frac{q_i}{2m_i}
  \left[ \vp_i\cdot\ovA(\vr_i) + \ovA(\vr_i)\cdot\vp_i \right]
\nonumber \\ & \quad
+ \sum_i \frac{{q_i}^2}{2m_i}\ovA^2(\vr_i).
\label{eq:rep-kinetic-cp} 
\end{align}
The first term is the kinetic energy without the radiation contribution,
and the other terms represent the interaction
between the charged particles and the radiation field.
Here, using the variables defined in Eqs.~\eqref{eq:def-I-micro} and 
\eqref{eq:def-N-micro}, 
we can rewrite Hamiltonian \eqref{eq:H-cp} as
\begin{align} \label{eq:H-cp2} 
\oH & = \oHem
+ \sum_i \left[ \frac{1}{2m_i} {\vp_i}^2 + V(\vr_i) \right]
+ \half \int\dd\vr\ \phicp \rhocp
\nonumber \\ & \quad
+ \int\dd\vr\ \phibg \rhocp
- \int\dd\vr \left[ \ovIcp \cdot \ovA
                  - \frac{1}{2}\oNcp\ovA^2 \right]
\end{align}
Expanding these terms with field operator \eqref{eq:def-ofld}, 
we obtain the first tree terms of interaction Hamiltonian
\eqref{eq:rep-oHint} from the exciton-associated components 
of the last three terms of Eq.~\eqref{eq:H-cp2}, i.e.,
the terms proportional to $\opcld_v\opcl_c$ or $\opcld_c\opcl_v$
but not to $\opcld_c\opcl_{c'}$ or $\opcld_v\opcl_{v'}$,
which are negligible under the weak excitation regime.
On the other hand, as mentioned in Sec.~\ref{sec:hamilt}, 
we put the exchange interaction between electrons and holes into $\oHint$.
It is obtained by expanding the fourth term of \eqref{eq:H-cp2}:
\begin{align}&
\half \int\dd\vr\ \ophicp \orhocp \rightarrow \cdots
\nonumber \\ & +
\sum_{c,v,c',v'} \opcld_c \opcl_v \opcld_{v'} \opcl_{c'}
\int\dd\vr\dd\vr'\ \frac{e^2\wvfpcl_c^*(\vr) \wvfpcl_v(\vr) 
  \wvfpcl_{v'}^*(\vr') \wvfpcl_{c'}(\vr')}{4\pi\diez|\vr-\vr'|}.
\label{eq:exp-Coulomb-cp} 
\end{align}
When we use the RWA and assume commutation relations \eqref{eq:[oex,oex]}
of the exciton operators,
we can find that the Coulomb interaction between the excitonic charges 
themselves
\begin{equation}
\frac{1}{2} \int\dd\vr\dd\vr'\ 
  \frac{\orhoex(\vr)\orhoex(\vr')}{4\pi\diez|\vr-\vr'|}
\end{equation}
gives exchange expression \eqref{eq:exp-Coulomb-cp}
with a constant energy term
by expanding $\orhoex(\vr)$ using Eq.~\eqref{eq:rep-orhoex-opcl}.
While all the other terms ($\cdots$) in Eq.~\eqref{eq:exp-Coulomb-cp}
and the second and third terms of Eq.~\eqref{eq:H-cp2}
belong to $\oHmat$,
instead of discussing them in detail,
we treat the matter Hamiltonian as Eq.~\eqref{eq:rep-oHmat}
for a simple description of the linear optical process of excitons
with nonradiative relaxation.
}
{
\def\longi#1{\left[#1\right]_{\text{L}}}
\def\trans#1{\left[#1\right]_{\text{T}}}
\def\wz{\omega_0}
\def\wzt{\tilde{\omega}_0}

\def\vB{\mathbf{B}}
\def\vC{\mathbf{C}}

\def\ofld{\psi}
\def\ofldd{\psi^{\dagger}}
\def\opcl{a}
\def\opcld{a^{\dagger}}
\def\wvfpcl{\varphi}

\def\ovJexT{\mathbf{J}_{\text{exT}}}
\def\ovJexL{\mathbf{J}_{\text{exL}}}
\def\ovPi{\mathbf{\Pi}}
\def\ovX{\mathbf{X}}
\def\ovP{\mathbf{P}}
\def\ovQ{\mathbf{Q}}
\def\ovY{\mathbf{Y}}
\def\oOmega{\Omega}

\def\ovJcp{\mathbf{J}_{\text{cp}}}
\def\ovIcp{\mathbf{I}_{\text{cp}}}
\def\ovPcp{\mathbf{P}_{\text{cp}}}
\def\oNcp{N_{\text{cp}}}
\def\orhocp{\rho_{\text{cp}}}
\def\ophicp{\phi_{\text{cp}}}

\def\dovJex{\dot{\mathbf{J}}_{\text{ex}}}
\def\dovPexL{\dot{\mathbf{P}}_{\text{exL}}}
\def\dorhoex{\dot{\rho}_{\text{ex}}}
\def\dophiex{\dot{\phi}_{\text{ex}}}

\def\hvJex{\hat{\mathbf{J}}_{\text{ex}}}
\def\hOmega{\hat{\Omega}}

\def\bchi{\bar{\chi}}
\def\bdie{\bar{\epsilon}}
\def\bvA{\bar{\mathbf{A}}}
\def\bvE{\bar{\mathbf{E}}}
\def\bvX{\bar{\mathbf{X}}}
\def\bvQ{\bar{\mathbf{Q}}}
\def\bvJ{\bar{\mathbf{J}}}
\def\bvJex{\bar{\mathbf{J}}_{\text{ex}}}
\def\bphiex{\bar{\phi}_{\text{ex}}}
\def\bOmega{\bar{\Omega}}

\def\cvE{\check{\mathbf{E}}}
\def\cvX{\check{\mathbf{X}}}
\def\cvJ{\check{\mathbf{J}}}
\def\cvJex{\check{\mathbf{J}}_{\text{ex}}}
\def\cphiex{\check{\phi}_{\text{ex}}}
\def\cOmega{\check{\Omega}}

\section{\label{app:qmaxwell}Extension of Maxwell wave equation}
Here, we extend the Maxwell wave equation discussed by SW
to enable the consideration of
the exciton-induced polarization with nonlocal susceptibility.
We derive the Heisenberg equations of the system variables
in Sec.~\ref{sec:eqmotion-AXY},
calculate their Laplace-transform in Sec.~\ref{sec:Laplace},
and provide the Maxwell wave equation for the electric field 
in Sec.~\ref{sec:q-Maxwell}.

\subsection{\label{sec:eqmotion-AXY}Heisenberg equations}
We derive the Heisenberg equations of 
the system variables and momenta from background Hamiltonian 
\eqref{eq:def-Hem} and interaction terms \eqref{eq:rep-oHint}.
The commutation relations of the variables are
\begin{align}
\left[ \ovA(\vr), \ovPi(\vr') \right] & = \ii\hbar\ \mdeltaT(\vr-\vr'),
\label{eq:[ovA,ovPi]} \\ 
\left[ \ovX(\vr), \ovP(\vr') \right] & = \ii\hbar\ \delta(\vr-\vr')\ \munit, \\
\left[ \ovY_{\omega}(\vr), \ovQ_{\omega'}(\vr') \right]
& = \ii\hbar\ \delta(\omega-\omega')\ \delta(\vr-\vr')\ \munit,
\end{align}
where
\begin{equation}\label{eq:def-deltaT}
\mdeltaT(\vr-\vr')
\equiv \munit\ \delta(\vr-\vr') + \frac{\vnabla'\vnabla'}{4\pi|\vr-\vr'|}
\end{equation}
is the Dirac delta function extracting the transverse component.
We obtain the equations of the radiation field as
\begin{align}
\dot{\ovA} & = \frac{1}{\diez}\ovPi,
\label{eq:Heisen-A} \\ 
\dot{\ovPi} & = \frac{1}{\muz}\Lapl\ovA
  - \trans{\frac{\alpha}{\rho}(\ovP+\alpha\ovA)}+ \ovJexT,
\label{eq:Heisen-Pi}
\end{align}
where 
\begin{equation}
\ovJexT(\vr) \equiv \int\dd\vr\ \mdeltaT(\vr-\vr') \cdot \ovJex(\vr')
\end{equation}
is the transverse component of the current density \eqref{eq:rep-ovJex}.
The equations of the polarizable oscillators are
\begin{align}
\dot{\ovX} & = \frac{1}{\rho}(\ovP+\alpha\ovA),
\label{eq:Heisen-X} \\ 
\dot{\ovP} & = - \rho\wzt{}^2\ovX - \frac{\alpha}{\diez}\longi{\alpha\ovX}
  - \frac{1}{\rho}\int_0^{\infty}\dd\omega\ v_{\omega}\ovQ_{\omega}
  + \alpha\grad\ophiex,
\label{eq:Heisen-P} 
\end{align}
and those of the reservoir oscillators are obtained as
\begin{align}
\dot{\ovY}_{\omega} & = \frac{1}{\rho}(\ovQ_{\omega}+v_{\omega}\ovX),
\label{eq:Heisen-Y} \\ 
\dot{\ovQ}_{\omega} & = - \rho\omega^2\ovY_{\omega}.
\label{eq:Heisen-Q} 
\end{align}

From Eqs.~\eqref{eq:Heisen-A}, \eqref{eq:Heisen-Pi}, and \eqref{eq:Heisen-X},
we obtain the Maxwell wave equation for the vector potential
\begin{equation}\label{eq:Maxwell-A-time}
\Lapl\ovA - \frac{1}{c^2}\ddot{\ovA}
= \muz\trans{\alpha\dot{\ovX}}- \muz\ovJexT,
\end{equation}
which has the transverse component of the excitonic current density
compared to the same kind of equation in Ref.~\onlinecite{suttorp04}.
Using a relation between the longitudinal components of excitonic variables
\begin{equation} \label{eq:rep-ovJexL} 
\ovJexL(\vr) = \dovPexL(\vr) = \diez\grad\dophiex(\vr)
\end{equation}
and that for the polarizable oscillators \eqref{eq:plz-L-bg}, 
we can rewrite Eq.~\eqref{eq:Maxwell-A-time}
as a wave equation for electric field \eqref{eq:rep-ovE}:
\begin{equation}\label{eq:Maxwell-E-time}
\rot\rot\ovE + \frac{1}{c^2}\ddot{\ovE}= \muz\alpha\ddot{\ovX}- \muz\dovJex.
\end{equation}
On the other hand, from Eqs.~\eqref{eq:Heisen-X} and \eqref{eq:Heisen-P},
we obtain the differential equation of the polarizable oscillators:
\begin{equation}\label{eq:wave-X}
\rho\ddot{\ovX}+ \rho\wzt{}^2\ovX
= \alpha\dot{\ovA}- \frac{\alpha}{\diez}\longi{\alpha\ovX}+ \alpha\grad\ophiex
- \frac{1}{\rho}\int_0^{\infty}\dd\omega\ v_{\omega}\ovQ_{\omega}.
\end{equation}

\subsection{\label{sec:Laplace}Laplace transform}
Next, we rewrite the equations of motion derived
in the previous section to those for the forward Laplace transform
of the variables
\begin{subequations} \label{eq:def-Laplace} 
\begin{equation}\label{eq:def-forward-Laplace}
\bOmega(p) \equiv \int_0^{\infty}\dd t\ \ee^{-pt}\oOmega(t),
\end{equation}
and for the backward Laplace transform
\begin{equation}\label{eq:def-backward-Laplace}
\cOmega(p) \equiv \int_0^{\infty}\dd t\ \ee^{-pt}\oOmega(-t).
\end{equation}
\end{subequations}
From these motion equations, we derive the ones
for positive-frequency Fourier transform
\begin{equation}\label{eq:def-Fourier}
\hOmega^+(\omega)
= \frac{1}{2\pi}
  \left[ \bOmega(-\ii\omega+\delta) + \cOmega(\ii\omega+\delta) \right].
\end{equation}

From the forward Laplace transform of Eqs.~\eqref{eq:Heisen-Y} 
and \eqref{eq:Heisen-Q} for the reservoir oscillators, 
we obtain
\begin{align}
\bvQ_{\omega}(p)
& = - \frac{\omega^2}{p^2+\omega^2}v_{\omega}\bvX(p)
\nonumber \\ & \quad
 + \frac{1}{p^2+\omega^2}
  \left[ p \ovQ_{\omega}(0) - \rho\omega^2\ovY_{\omega}(0) \right],
\end{align}
and, by using Eq.~\eqref{eq:Heisen-X} at $t = 0$,
that for the polarizable oscillators \eqref{eq:wave-X} becomes
\begin{align}
&   (p^2+\wzt{}^2)\bvX(p) \nonumber \\
& = \frac{\alpha}{\rho}\left\{ p\bvA(p) - \frac{1}{\diez}\longi{\alpha\bvX(p)}
                            + \grad\bphiex(p) \right\} \nonumber \\
& \quad - \frac{1}{\rho^2}\int_0^{\infty}\dd\omega\ v_{\omega}\bvQ_{\omega}(p)
  + p\ovX(0) + \frac{1}{\rho}\ovP(0).
\end{align}
Substituting the former into the latter and using the forward Laplace-transform
of the electric field
\begin{equation}
\bvE(p) = - p \bvA(p) + \frac{1}{\diez}\longi{\alpha\bvX(p)}
  - \grad\bphiex(p) + \ovA(0),
\end{equation}
we obtain the same equation as Eq.~(20) of Ref.~\onlinecite{suttorp04}:
\begin{align}&
\bvX(p)
= - \frac{\diez}{\alpha}\bchi(p)\bvE(p)
  + \frac{\diez}{\alpha^2}\bchi(p) \biggl\{
    \alpha\ovA(0) + \rho p\ovX(0)
\nonumber \\ & \quad
   + \ovP(0) + \int_0^{\infty}\dd\omega\ \frac{v_{\omega}}{p^2+\omega^2}
    \left[\omega^2\ovY_{\omega}(0)-\frac{p}{\rho}\ovQ_{\omega}(0)\right]
  \biggr\},
\end{align}
where
\begin{equation}
\bchi(p) = \frac{\alpha^2}{\diez\rho}\left[
  p^2 + \wzt{}^2 - \frac{1}{\rho^2}\int_0^{\infty}\dd\omega\
  \frac{\omega^2{v_{\omega}}^2}{p^2+\omega^2}
\right]^{-1}
\end{equation}
is the background susceptibility.

Next, we derive the forward Laplace transform of Maxwell wave equation 
\eqref{eq:Maxwell-E-time} for the electric field as
\begin{align}\label{eq:forward-Maxwell-1}
& \rot\rot\bvE(p) + \frac{p^2}{c^2}\bvE(p) - \mu\alpha p^2\bvX(p)
\nonumber \\ &
= - \muz p\bvJex(p) + \muz\ovJex(0)
+ \frac{1}{c^2}\dot{\ovE}(0)+ \frac{p}{c^2}\ovE(0) 
\nonumber \\ & \quad
- \muz\alpha\dot{\ovX}(0) - \muz\alpha p\ovX(0).
\end{align}
Here, using the Eqs.~\eqref{eq:rep-ovE}, \eqref{eq:plz-L-bg},
\eqref{eq:Heisen-A}, \eqref{eq:Heisen-X}, \eqref{eq:Maxwell-A-time},
and \eqref{eq:rep-ovJexL}, we obtain the relations between the variables at
$t = 0$ as
\begin{equation}
\dot{\ovX}(0) = \frac{\alpha}{\rho}\ovA(0) + \frac{1}{\rho}\ovP(0),
\end{equation}
\begin{equation}
\dot{\ovE}(0) = c^2\rot\rot\ovA(0) + \frac{\alpha^2}{\diez\rho}\ovA(0)
  + \frac{\alpha}{\diez\rho}\ovP(0) - \frac{1}{\diez}\ovJex(0),
\end{equation}
\begin{equation}
\ovE(0) = - \frac{1}{\diez}\ovPi(0) + \frac{1}{\diez}\longi{\alpha\ovX(0)}
  - \grad\ophiex(0).
\end{equation}
Using these relations, we can rewrite the Maxwell wave equation
\eqref{eq:forward-Maxwell-1} to
\begin{align}&
\rot\rot\bvE(p) + \frac{p^2}{c^2}\bdie(p)\bvE(p)
\nonumber \\ &
= - \muz p\bvJ(p) - \muz p\bvJex(p) - \frac{p}{c^2}\grad\ophiex(0),
\label{eq:forward-Maxwell}
\end{align}
where $\bdie(p) =1+ \bchi(p)$ is the background dielectric function,
and the operator
\begin{align}
& \bvJ(p)
  = - \frac{1}{\muz p}\rot\rot\ovA(0) - \diez p\bchi(p)\ovA(0) + \ovPi(0)
\nonumber \\ & \quad
  + \alpha\left[1-\frac{\diez\rho}{\alpha^2}p^2\bchi(p)\right]\ovX(0)
  - \longi{\alpha\ovX(0)}- \frac{\diez}{\alpha}p\bchi(p)\ovP(0)
\nonumber \\ & \quad
  - \frac{\diez}{\alpha}p\bchi(p)\int_0^{\infty}\dd\omega\
    \frac{v_{\omega}}{p^2+\omega^2}
    \left[ \omega^2\ovY_{\omega}(0) - \frac{p}{\rho}\ovQ_{\omega}(0) \right]
\end{align}
is the same one shown in Eq.~(27) of Ref.~\onlinecite{suttorp04}.
This operator depends only on the background system variables and momenta
of Ref.~\onlinecite{suttorp04}.

On the other hand, the backward Laplace transform of the
Maxwell wave equation is obtained as
\begin{align}&
\rot\rot\cvE(p) + \frac{p^2}{c^2}\bdie(p)\cvE(p)
\nonumber \\ &
= \muz p\cvJ(p) + \muz p\cvJex(p) - \frac{p}{c^2}\grad\ophiex(0),
\label{eq:backward-Maxwell}
\end{align}
where the operator on the RHS
\begin{align}
& \cvJ(p)
= - \frac{1}{\muz p}\rot\rot\ovA(0) - \diez p\bchi(p)\ovA(0) - \ovPi(0)
\nonumber \\ & \quad
  - \alpha\left[1-\frac{\diez\rho}{\alpha^2}p^2\bchi(p)\right]\ovX(0)
  + \longi{\alpha\ovX(0)}- \frac{\diez}{\alpha}p\bchi(p)\ovP(0)
\nonumber \\ & \quad
  - \frac{\diez}{\alpha}p\bchi(p)\int_0^{\infty}\dd\omega\
    \frac{v_{\omega}}{p^2+\omega^2}
    \left[ \omega^2\ovY_{\omega}(0) + \frac{p}{\rho}\ovQ_{\omega}(0) \right]
\label{eq:def-cvJ} 
\end{align}
is also independent from the variables associated with excitons.

\subsection{\label{sec:q-Maxwell}Fourier transform}
From the forward and backward Laplace transforms 
\eqref{eq:forward-Maxwell} and \eqref{eq:backward-Maxwell}
of the Maxwell wave equation,
we obtain that for the positive-frequency Fourier component
of the electric field as
\begin{align}&
\rot\rot\hvE^+(\vr,\omega) - \frac{\omega^2}{c^2}\dieb(\vr,\omega)
  \hvE^+(\vr,\omega)
\nonumber \\ &
= \ii\muz\omega \left[ \hvJz(\vr,\omega) + \hvJex^+(\vr,\omega) \right],
\label{eq:Maxwell-Jz-Jex}
\end{align}
where $\dieb(\vr,\omega) = \bdie(\vr,-\ii\omega+\delta)$ is
the background dielectric function.
The noise current density operator in our system is written as
\begin{align}&
\hvJz(\vr,\omega)
= \hvJ(\vr,\omega)
- \frac{\ii\omega^2}{\pi c^2} \Im[\dieb(\vr,\omega)]
\nonumber \\ & \quad \times
  \int\dd\vr'\ \mG^*(\vr,\vr',\omega)
  \cdot \left[ \cvJex(\vr',\ii\omega+\delta) - \diez\grad\ophiex(\vr',0) \right],
\label{eq:def-Jz}
\end{align}
where $\mG(\vr,\vr,\omega)$ is Green's tensor satisfying 
Eq.~\eqref{eq:satisfied-mG}, and
\begin{align}&
\hvJ(\vr,\omega)
= \frac{1}{2\pi}
  \left[ \bvJ(\vr,-\ii\omega+\delta) + \cvJ(\vr,\ii\omega+\delta) \right]
\nonumber \\ & \quad
- \frac{\ii\omega^2}{\pi c^2} \Im[\dieb(\vr,\omega)]
  \int\dd\vr'\ \mG^*(\vr,\vr',\omega) \cdot \cvJ(\vr',\ii\omega+\delta)
\label{eq:def-J-Fourier}
\end{align}
is the same operator shown in Eq.~(45) of Ref.~\onlinecite{suttorp04}
satisfying the commutation relations \eqref{eq:[hvJ,hvJ]}.
By using the relation
\begin{equation}
\hvJex(\vr, \omega) = - \ii\omega\hvPex(\vr, \omega),
\end{equation}
we can rewrite Eq.~\eqref{eq:Maxwell-Jz-Jex} to \eqref{eq:Maxwell-E-Jz-Pex}.
}
{
\def\longi#1{\left[#1\right]_{\text{L}}}
\def\trans#1{\left[#1\right]_{\text{T}}}

\def\mX{\bm{\mathsf{X}}}

\def\ophi{\phi}
\def\ovPexL{\mathbf{P}_{\text{exL}}}
\def\ovPexT{\mathbf{P}_{\text{exT}}}

\def\hvJex{\hat{\mathbf{J}}_{\text{ex}}}
\def\hsrc{\hat{\beta}}
\def\hsrcd{\hat{\beta}^{\dagger}}
\def\hsrca{\hat{\beta}^{\text{abs}}}

\def\hvA{\hat{\mathbf{A}}}
\def\hphi{\hat{\phi}}

\def\bex{\bar{b}}
\def\bexd{\bar{b}^{\dagger}}
\def\bsrc{\bar{\beta}}
\def\bsrca{\bar{\beta}^{\text{abs}}}
\def\bres{\bar{d}}
\def\bresd{\bar{d}^{\dagger}}
\def\blg{\bar{D}}
\def\bphi{\bar{\phi}}
\def\bphiex{\bar{\phi}_{\text{ex}}}
\def\bvJex{\bar{\mathbf{J}}_{\text{ex}}}
\def\bvPex{\bar{\mathbf{P}}_{\text{ex}}}
\def\bvA{\bar{\mathbf{A}}}
\def\bvE{\bar{\mathbf{E}}}
\def\bvJ{\bar{\mathbf{J}}}

\def\cex{\check{b}}
\def\cexd{\check{b}^{\dagger}}
\def\csrc{\check{\beta}}
\def\csrca{\check{\beta}^{\text{abs}}}
\def\cres{\check{d}}
\def\cresd{\check{d}^{\dagger}}
\def\clg{\check{D}}
\def\cphi{\check{\phi}}
\def\cphiex{\check{\phi}_{\text{ex}}}
\def\cvJex{\check{\mathbf{J}}_{\text{ex}}}
\def\cvPex{\check{\mathbf{P}}_{\text{ex}}}
\def\cvA{\check{\mathbf{A}}}
\def\cvE{\check{\mathbf{E}}}
\def\cvJ{\check{\mathbf{J}}}

\section{\label{app:qlinopt}Evaluation of commutators}
Although we derive the Maxwell wave equation considering
the excitons in App.~\ref{app:qmaxwell},
there exists a problem in deriving the commutation relations 
of noise current density $\hvJz(\vr,\omega)$ 
and in describing the optical processes of excitons.
These are the subjects of this appendiex.
In Sec.~\ref{sec:eq-lin-opt-ex}, we derive the self-consistent equations
for the Laplace-transformed operators,
and we evaluate the commutators
of the noise current density in Sec.~\ref{sec:commut-Ez}.
In Sec.~\ref{sec:with-abs}, we consider nonradiative relaxation of excitons.

\subsection{\label{sec:eq-lin-opt-ex}Laplace transform of self-consistent 
equations}
From matter Hamiltonian \eqref{eq:rep-oHmat} 
and interaction Hamiltonian \eqref{eq:rep-oHint},
neglecting the radiation contribution of the current density 
$\oNex(\vr)\ovA^2(\vr)/2$ under weak excitation,
we obtain the Heisenberg equation of excitons as
\begin{align} \label{eq:motion-ex-time} 
& \ii\hbar\ddt{}\oex_{\mu}(t)
\nonumber \\ &
= \hbar\wex_{\mu}\oex_{\mu}(t)
- \int\dd\vr \left[ \vdimI_{\mu}^*(\vr) \cdot \ovA(\vr,t)
  - \rho_{\mu}^*(\vr) \ophi(\vr,t) \right]
\nonumber \\ & \quad
+ \int_0^{\infty}\dd\wres
  \left[ \ccxr_{\mu}(\wres) \ores_{\mu}(\wres,t)
       + \ccxr^*_{\mu}(\wres) \oresd_{\mu}(\wres,t) \right],
\end{align}
where $\ophi(\vr) \equiv \ophibg(\vr) + \ophiex(\vr)$ is
the complete Coulomb potential.
In this section, we neglect nonradiative relaxation
and assume $\omega \sim \wex_{\mu}$.
Using the relations between $\vdimI_{\mu}(\vr)$, $\rho_{\mu}(\vr)$, 
and $\vdimP_{\mu}(\vr)$ (Eqs.~\eqref{eq:vdimI=w*vdimP} and 
\eqref{eq:rho=div*vdimP}) and the Laplace transform of the electric field
\begin{subequations}
\begin{align}
\bvE(\vr,p) & = -p\bvA(\vr,p) - \grad\bphi(\vr,p) + \ovA(\vr,0), \\
\cvE(\vr,p) & = p\cvA(\vr,p) - \grad\cphi(\vr,p) - \ovA(\vr,0), 
\end{align}
\end{subequations}
the forward and backward Laplace transforms of Eq.~\eqref{eq:motion-ex-time} 
are respectively derived as
\begin{subequations} \label{eq:Laplace-ex-2} 
\begin{align}& 
(\hbar\wex_{\mu}-\hbar\omega-\ii\delta)\ \bex_{\mu}(-\ii\omega+\delta)
= - \ii\hbar\oex_{\mu}(0) 
\nonumber \\ & \quad
+ \int\dd\vr\ \vdimP_{\mu}^*(\vr) \cdot
  \left[ \bvE(\vr,-\ii\omega+\delta) - \ovA(\vr,0) \right],
\end{align}
\begin{align}& 
(\hbar\wex_{\mu}-\hbar\omega+\ii\delta)\ \cex_{\mu}(\ii\omega+\delta)
= \ii\hbar\oex_{\mu}(0) 
\nonumber \\ & \quad
+ \int\dd\vr\ \vdimP_{\mu}^*(\vr) \cdot
    \left[ \cvE(\vr, \ii\omega+\delta) + \ovA(\vr,0) \right].
\end{align}
\end{subequations}
Adding these two equations, we obtain Eq.~\eqref{eq:motion-hexone},
the Fourier transform of the excitons' motion equation.
On the other hand,
using the relations
\begin{equation} \label{eq:ophiex=ovPexL} 
\diez\grad\ophiex(\vr) = \ovPexL(\vr),
\end{equation}
and
\begin{subequations} \label{eq:Lapl-vJex} 
\begin{align}
\bvJex(\vr,p) & = p \bvPex(\vr,p) - \ovPex(\vr,0), \\
\cvJex(\vr,p) & = - p \cvPex(\vr,p) + \ovPex(\vr,0),
\end{align}
\end{subequations}
the Laplace transforms \eqref{eq:forward-Maxwell} 
and \eqref{eq:backward-Maxwell} of the Maxwell wave equation are rewritten as
\begin{subequations} \label{eq:Laplace-Maxwell-3} 
\begin{align}
& \rot\rot\bvE(\vr,-\ii\omega+\delta)
  - \frac{\omega^2}{c^2}\dieb(\vr, \omega)\bvE(\vr,-\ii\omega+\delta)
\nonumber \\ &
= \ii\muz\omega \left[ \bvJ(\vr,-\ii\omega+\delta) - \ovPexT(\vr,0) \right]
\nonumber \\ & \quad
+ \muz\omega^2\bvPex(\vr,-\ii\omega+\delta),
\end{align}
\begin{align}
& \rot\rot\cvE(\vr, \ii\omega+\delta)
  - \frac{\omega^2}{c^2}\dieb^*(\vr, \omega)\cvE(\vr, \ii\omega+\delta)
\nonumber \\ & 
= \ii\muz\omega \left[ \cvJ(\vr, \ii\omega+\delta) + \ovPexT(\vr,0) \right]
\nonumber \\ & \quad
  + \muz\omega^2\cvPex(\vr, \ii\omega+\delta).
\end{align}
\end{subequations}
Here, under the RWA, the Laplace transforms of the polarization density 
operator can be written as
\begin{subequations} \label{eq:rwa-Laplace-P} 
\begin{align}
\bvPex(\vr,-\ii\omega+\delta)
& \sim \sum_{\mu}\vdimP_{\mu}(\vr) \bex_{\mu}(-\ii\omega+\delta), \\
\cvPex(\vr,\ii\omega+\delta)
& \sim \sum_{\mu}\vdimP_{\mu}(\vr) \cex_{\mu}(\ii\omega+\delta)
\label{eq:rwa-backward-P} 
\end{align}
\end{subequations}
in the same manner as its Fourier transform \eqref{eq:Pex=P*ex-rwa}.
Substituting Maxwell wave equations \eqref{eq:Laplace-Maxwell-3}
in motion equations \eqref{eq:Laplace-ex-2} of excitons,
we obtain the self-consistent equation set
for the Laplace transformed operators:
\begin{subequations} \label{eq:Laplace-self-consist} 
\begin{align}
& \sum_{\mu'}\left[ (\hbar\wex_{\mu}-\hbar\omega)\delta_{\mu, \mu'}
  + \rct_{\mu, \mu'}(\omega) \right] \bex_{\mu'}(-\ii\omega+\delta)
\nonumber \\ &
= \bsrc_{\mu}(-\ii\omega+\delta),
\end{align}
\begin{align}
& \sum_{\mu'}\left[ (\hbar\wex_{\mu}-\hbar\omega)\delta_{\mu, \mu'}
  + \rct_{\mu', \mu}^*(\omega) \right] \cex_{\mu'}(\ii\omega+\delta)
\nonumber \\ &
= \csrc_{\mu}(\ii\omega+\delta), \label{eq:backward-self-consist} 
\end{align}
\end{subequations}
where the operators on the RHS are
\begin{subequations} \label{eq:Laplace-src-ex} 
\begin{align}
\bsrc_{\mu}(-\ii\omega+\delta)
& \equiv \ii\muz\omega \int\dd\vr\int\dd\vr'\ \vdimP_{\mu}^*(\vr)
  \cdot \mG(\vr, \vr', \omega)
\nonumber \\ & \quad
  \cdot \left[ \bvJ(\vr',-\ii\omega+\delta) - \ovPexT(\vr',0) \right]
\nonumber \\ & \quad
- \ii\hbar\oex_{\mu}(0) - \int\dd\vr\ \vdimP_{\mu}^*(\vr) \cdot \ovA(\vr,0),
\end{align}
\begin{align} \label{eq:backward-src-ex} 
\csrc_{\mu}(\ii\omega+\delta)
& \equiv \ii\muz\omega \int\dd\vr\int\dd\vr'\ \vdimP_{\mu}^*(\vr)
  \cdot \mG^*(\vr, \vr', \omega)
\nonumber \\ & \quad
  \cdot \left[ \cvJ(\vr', \ii\omega+\delta) + \ovPexT(\vr',0) \right]
\nonumber \\ & \quad
+ \ii\hbar\oex_{\mu}(0) + \int\dd\vr\ \vdimP_{\mu}^*(\vr) \cdot \ovA(\vr,0).
\end{align}
\end{subequations}
In order to derive \eqref{eq:backward-self-consist}, 
we used the representation
\begin{equation}
\rct_{\mu', \mu}^*(\omega)
= - \mu\omega^2 \int\dd\vr\int\dd\vr'\ \vdimP_{\mu}^*(\vr)
  \cdot \mG^*(\vr, \vr', \omega) \cdot \vdimP_{\mu'}(\vr'),
\end{equation}
which comes from definition \eqref{eq:def-rct} and the reciprocity relation
\begin{equation} \label{eq:G-exch-pos} 
G_{ij}(\vr, \vr', \omega) = G_{ji}(\vr', \vr, \omega)
\end{equation}
shown in Eq.~(1.53) of Ref.~\onlinecite{knoll01}.
Adding self-consistent equation sets \eqref{eq:Laplace-self-consist},
we obtain the one for the Fourier transform of the exciton operators as
\begin{equation}\label{eq:self-consist-1} 
\sum_{\mu'}\left[ (\hbar\wex_{\mu}-\hbar\omega)\delta_{\mu, \mu'}
  + \rct_{\mu, \mu'}(\omega) \right] \hex_{\mu'}(\omega)
= \hsrc_{\mu}(\omega),
\end{equation}
where the operator on the RHS is
\begin{align} &
\hsrc_{\mu}(\omega)
\equiv \frac{1}{2\pi}\bsrc_{\mu}(-\ii\omega+\delta)
+ \frac{1}{2\pi}\csrc_{\mu}(\ii\omega+\delta)
\nonumber \\ & \quad
+ \frac{1}{2\pi}\sum_{\mu'}
  \left[ \rct_{\mu, \mu'}(\omega) - \rct_{\mu', \mu}^*(\omega)\right]
  \cex_{\mu'}(\ii\omega+\delta).
\label{eq:src-ex-1}
\end{align}
Now, we verify that this operator is equivalent to the RHS of 
Eq.~\eqref{eq:self-consist-S}.
Using Eqs.~\eqref{eq:def-Jz}, \eqref{eq:ophiex=ovPexL}, \eqref{eq:Lapl-vJex},
and \eqref{eq:G(e-e)G=G-G},
the background electric field \eqref{eq:def-hvEz-hvJz} is rewritten as
\begin{align} \label{eq:Ez-repre-Pex-1} 
\hvEz^+(\vr, \omega)
& = \ii\muz\omega \int\dd\vr'\ \mG(\vr, \vr', \omega) \cdot \hvJ(\vr', \omega)
\nonumber \\ & \quad
- \frac{\muz}{2\pi}\int\dd\vr'\
  \left[ \mG(\vr, \vr', \omega) - \mG^*(\vr, \vr', \omega) \right]
\nonumber \\ & \quad 
  \cdot \left[ \omega^2 \cvPex(\vr', \ii\omega+\delta)
       + \ii\omega \ovPexT(\vr',0) \right].
\end{align}
Since, from Eqs.~\eqref{eq:def-J-Fourier} and \eqref{eq:G(e-e)G=G-G},
we can rewrite the first term as
\begin{align}
& \int\dd\vr'\ \mG(\vr, \vr', \omega) \cdot \hvJ(\vr', \omega)
\nonumber \\ &
= \frac{1}{2\pi}\int\dd\vr'\
  \mG(\vr, \vr', \omega) \cdot \bvJ(\vr',-\ii\omega+\delta)
\nonumber \\ & \quad
+ \frac{1}{2\pi}\int\dd\vr'\
  \mG^*(\vr, \vr', \omega) \cdot \cvJ(\vr', \ii\omega+\delta),
\end{align}
background field \eqref{eq:Ez-repre-Pex-1} can be rewritten as
\begin{widetext}
\begin{align}\label{eq:Ez-repre-Pex-2}
\hvEz^+(\vr, \omega)
& = \frac{\ii\muz\omega}{2\pi}\int\dd\vr'\ \mG(\vr, \vr', \omega) \cdot
    \left[ \bvJ(\vr',-\ii\omega+\delta) - \ovPexT(\vr',0) \right]
\nonumber \\ & \quad
+ \frac{\ii\muz\omega}{2\pi}\int\dd\vr'\ \mG^*(\vr, \vr', \omega) \cdot
  \left[ \cvJ(\vr', \ii\omega+\delta) + \ovPexT(\vr',0) \right]
\nonumber \\ & \quad
- \frac{\muz\omega^2}{2\pi}\int\dd\vr'\
  \left[ \mG(\vr, \vr', \omega) - \mG^*(\vr, \vr', \omega) \right]
  \cdot \cvPex(\vr', \ii\omega+\delta),
\end{align}
and the RHS of Eq.~\eqref{eq:self-consist-S} becomes
\begin{align}
\int\dd\vr\ \vdimP_{\mu}^*(\vr) \cdot \hvEz^+(\vr, \omega)
& = \frac{\ii\muz\omega}{2\pi}\int\dd\vr\int\dd\vr'\ \vdimP_{\mu}^*(\vr)
  \cdot \mG(\vr, \vr', \omega) \cdot
  \left[ \bvJ(\vr',-\ii\omega+\delta) - \ovPexT(\vr',0) \right]
\nonumber \\ & \quad
+ \frac{\ii\muz\omega}{2\pi}\int\dd\vr\int\dd\vr'\ \vdimP_{\mu}^*(\vr)
  \cdot \mG^*(\vr, \vr', \omega) \cdot
  \left[ \cvJ(\vr', \ii\omega+\delta) + \ovPexT(\vr',0) \right]
\nonumber \\ & \quad
+ \frac{1}{2\pi}\sum_{\mu'}
  \left[ \rct_{\mu, \mu'}(\omega) - \rct_{\mu', \mu}^*(\omega) \right]
  \cex_{\mu'}(\ii\omega+\delta).
\end{align}
\end{widetext}
We can find that this is equivalent to \eqref{eq:src-ex-1}
by expanding $\bsrc_{\mu}(-\ii\omega+\delta)$ and 
$\csrc_{\mu}(\ii\omega+\delta)$ with Eqs.~\eqref{eq:Laplace-src-ex}.

\subsection{\label{sec:commut-Ez}Commutation relations}
In the representation of background field \eqref{eq:Ez-repre-Pex-1}
or \eqref{eq:Ez-repre-Pex-2}, the information of excitons during $t < 0$
is reflected via operator $\cvPex(\vr, \ii\omega+\delta)$.
In order to evaluate the commutation relations
of noise current density $\hvJz(\vr, \omega)$,
we first discuss the motion equations of excitons.

The exciton motion during $t < 0$ is described 
by Eq.~\eqref{eq:backward-self-consist} with \eqref{eq:backward-src-ex}.
Using $\mcrd(\omega)$, the inverse matrix of coefficient matrix
$\mS(\omega)$ defined in Eq.~\eqref{eq:def-element-mS},
we rewrite Eq.~\eqref{eq:backward-self-consist} to
\begin{equation} \label{eq:b=Wsrc} 
\cex_{\mu}(\ii\omega+\delta)
= \sum_{\mu'}\crd^*_{\mu', \mu}(\omega)\ \csrc_{\mu'}(\ii\omega+\delta).
\end{equation}
First of all, we evaluate the cummutation relations of
$\csrc_{\mu}(\ii\omega+\delta)$.
Since, from Eqs.~\eqref{eq:def-cvJ} and \eqref{eq:rep-ovPex-oex},
we obtain the relations 
\begin{equation}
\left[ \cvJ(\vr, \ii\omega+\delta), \ovA(\vr',0) \right]
= \ii\hbar\mdeltaT(\vr-\vr'),
\end{equation}
\begin{equation}
\left[ \ovPexT(\vr,0), \oex_{\mu}(0) \right] = - \trans{\vdimP_{\mu}^*(\vr)},
\end{equation}
the first two terms and the following terms of \eqref{eq:backward-src-ex}
are commutable as
\begin{align} \label{eq:[cJ+PT,b+PA]}
& \biggl[ \cvJ(\vr, \ii\omega+\delta) + \ovPexT(\vr,0),
\nonumber \\ & \quad \left.
       \ii\hbar\oex_{\mu}(0)
     + \int\dd\vr'\ \vdimP^*_{\mu}(\vr') \cdot \ovA(\vr',0) \right] = \vzero.
\end{align}
Then, since $\ovPexT(\vr,0)$ and $\ovA(\vr,0)$ are commutable with themselves,
we obtain the expression to be evaluated:
\begin{align}&
\left[ \csrc_{\mu}(\ii\omega+\delta),
       \dgg{\csrc_{\nu}(\ii\omega'+\delta')}\right]
= \hbar^2\delta_{\mu, \nu}
\nonumber \\ &
+ {\muz}^2\omega\omega' \int\dd\vr\int\dd\vs\int\dd\vs'\int\dd\vr'\
  \vdimP_{\mu}^*(\vr) \cdot \mG^*(\vr, \vs, \omega)
\nonumber \\ & \quad
  \cdot
  \left[ \cvJ(\vs, \ii\omega+\delta), \dgg{\cvJ(\vs', \ii\omega'+\delta')}
  \right]
  \cdot \mG(\vs', \vr', \omega') \cdot \vdimP_{\nu}(\vr').
\label{eq:backward-[src,srcd]} 
\end{align}
Using the relation shown in Eq.~(B1) of Ref.~\onlinecite{suttorp04}
\begin{align}
& \left[ \cvJ(\vs, \ii\omega+\delta),
  \dgg{\cvJ(\vs', \ii\omega'+\delta')}\right]
\nonumber \\ &
= \frac{\hbar(\omega+\omega')}{\muz\omega\omega'}
  \left[ \vnabla\vnabla - \munit \Lapl \right] \delta(\vs-\vs')
\nonumber \\ & \quad
+ \frac{\diez\hbar\omega\omega'}{\omega-\omega'-\ii\delta}
  \left[ \dieb^*(\vs, \omega) - \dieb(\vs, \omega') \right] \munit
  \delta(\vs-\vs'),
\end{align}
we obtain
\begin{align} \label{eq:G[J,Jd]G=(G-G)/(w-w)} 
& {\muz}^2\omega\omega'\int\dd\vs\int\dd\vs'\ \mG^*(\vr, \vs, \omega)
\nonumber \\ & \quad
\cdot \left[ \cvJ(\vs, \ii\omega+\delta),
      \dgg{\cvJ(\vs', \ii\omega'+\delta')}\right]
    \cdot \mG(\vs', \vr', \omega')
\nonumber \\ &
= \frac{\muz \hbar}{\omega-\omega'-\ii\delta}\left[
    \omega^2 \mG^*(\vr, \vr', \omega)
  - {\omega'}^2 \mG(\vr, \vr', \omega') \right].
\end{align}
From this relation, Eq.~\eqref{eq:backward-[src,srcd]} is evaluated as
\begin{align}
&  \left[ \csrc_{\mu}(\ii\omega+\delta),
   \dgg{\csrc_{\mu'}(\ii\omega'+\delta')}\right] \nonumber \\
& = \hbar^2\delta_{\mu, \mu'}+ \frac{\hbar}{\omega-\omega'-\ii\delta}
    \left[ \rct_{\mu, \mu'}(\omega') - \rct_{\mu', \mu}^*(\omega) \right] \\
& = \frac{\hbar}{\omega-\omega'-\ii\delta}
    \left[ S_{\mu, \mu'}(\omega') - S_{\mu', \mu}^*(\omega) \right].
\label{eq:[src,srcd]=S-S} 
\end{align}
We also obtain the relation
\begin{equation}\label{eq:[src,src]=0}
\left[ \csrc_{\mu}(\ii\omega+\delta),
       \csrc_{\mu'}(\ii\omega'+\delta') \right] = 0,
\end{equation}
because this commutator has a nonresonant denominator
compared to \eqref{eq:[src,srcd]=S-S}.
Therefore, from Eqs.~\eqref{eq:b=Wsrc}, \eqref{eq:[src,srcd]=S-S},
and \eqref{eq:[src,src]=0},
we obtain the commutation relations of the backward Laplace-transformed
exciton operators
\begin{subequations} \label{eq:[cex,cex]} 
\begin{align}
& \left[ \cex_{\mu}(\ii\omega+\delta),
  \dgg{\cex_{\mu'}(\ii\omega'+\delta')}\right]
\nonumber \\ & \quad
= \frac{\hbar}{\omega-\omega'-\ii\delta}
    \left[ \crd_{\mu', \mu}^*(\omega) - \crd_{\mu, \mu'}(\omega') \right],
\\ &
\left[ \cex_{\mu}(\ii\omega+\delta),
       \cex_{\mu'}(\ii\omega'+\delta') \right]
= 0.
\end{align}
\end{subequations}

On the other hand, using relations \eqref{eq:ophiex=ovPexL} and
\eqref{eq:Lapl-vJex}, noise current density $\hvJz(\vr,\omega)$ defined in
Eq.~\eqref{eq:def-Jz} is rewritten as
\begin{align}
\hvJz(\vr,\omega)
& = \hvJ(\vr,\omega) - \frac{\omega^3}{\pi c^2}\
    \Im[\dieb(\vr,\omega)]\int\dd\vr'\ \mG^*(\vr,\vr',\omega)
\nonumber \\ & \quad
    \cdot \left[ \cvPex(\vr',\ii\omega+\delta)
               + \frac{\ii}{\omega} \ovPexT(\vr',0) \right].
\label{eq:rep-hvJz-P} 
\end{align}
Then, the commutator with its Hermite conjugate is evaluated as
\begin{align}
\left[ \hvJz(\vr,\omega), \dgg{\hvJz(\vr',\omega')} \right]
& = \left[ \hvJ(\vr,\omega), \dgg{\hvJ(\vr',\omega')} \right]
\nonumber \\ & \quad
    + \mX_1 + \mX_2 + \mX_3,
\end{align}
where
\begin{align}
\mX_1
& \equiv - \frac{\omega^3}{\pi c^2}\ \Im[\dieb(\vr,\omega)]
\nonumber \\ & \quad \times
  \int\dd\vs\ \mG^*(\vr,\vs,\omega) \cdot
  \left[ \cvPex(\vs,\ii\omega+\delta), \dgg{\hvJ(\vr',\omega')} \right], \\
\mX_2
& \equiv - \frac{{\omega'}^3}{\pi c^2}\ \Im[\dieb(\vr',\omega')]
\nonumber \\ & \quad \times
  \int\dd\vs'\ 
  \left[ \hvJ(\vr,\omega), \dgg{\cvPex(\vs',\ii\omega'+\delta)} \right]
  \cdot \mG(\vs',\vr',\omega'), \\
\mX_3
& \equiv \frac{\omega^3{\omega'}^3}{\pi^2 c^4}\
  \Im[\dieb(\vr,\omega)]\ \Im[\dieb(\vr',\omega')]
\nonumber \\ & \quad \times
  \int\dd\vs\int\dd\vs'\ \mG^*(\vr,\vs,\omega)
\nonumber \\ & \quad
  \cdot
  \biggl[ \cvPex(\vs,\ii\omega+\delta) + \frac{\ii}{\omega}\ovPexT(\vs,0),
\nonumber \\ & \quad
  \dgg{\cvPex(\vs',\ii\omega'+\delta)} - \frac{\ii}{\omega'}\ovPexT(\vs',0)
  \biggr] \cdot \mG(\vs',\vr',\omega').
\end{align}
First, from Eqs.~\eqref{eq:Laplace-src-ex} and \eqref{eq:b=Wsrc},
the commutator appearing in $\mX_1$ becomes
\begin{align}
&   \left[ \cvPex(\vs,\ii\omega+\delta), \dgg{\hvJ(\vr',\omega')} \right]
\nonumber \\ &
= \sum_{\mu,\mu'} \vdimP_{\mu}(\vs)\ \crd_{\mu',\mu}^*(\omega)
  \int\dd\vs'\ \vdimP_{\mu'}^*(\vs')
\nonumber \\ & \quad \cdot
  \biggl[ \ii\muz\omega \int\dd\vs''\ \mG^*(\vs',\vs'',\omega)
    \cdot \cvJ(\vs'',\ii\omega+\delta) + \ovA(\vs',0),
\nonumber \\ & \quad
    \dgg{\hvJ(\vr',\omega')} \biggr].
\label{eq:[cvPex,hvJd]} 
\end{align}
Since $\hvJ(\vr',\omega')$ is defined as Eq.~\eqref{eq:def-J-Fourier},
what we must evaluate is 
\begin{align}
&   \left[ \cvJ(\vs'',\ii\omega+\delta), \dgg{\hvJ(\vr',\omega')} \right]
\nonumber \\ &
= \frac{1}{2\pi} \left[ \cvJ(\vs'',\ii\omega+\delta),
  \dgg{\bvJ(\vr',-\ii\omega'+\delta) + \cvJ(\vr',\ii\omega'+\delta)} \right]
\nonumber \\ & \quad
+ \frac{\ii{\omega'}^2}{\pi c^2}\ \Im[\dieb(\vr',\omega')] \int\dd\vr''
\nonumber \\ & \quad
 \left[ \cvJ(\vs'',\ii\omega+\delta),
    \dgg{\cvJ(\vr'',\ii\omega'+\delta)} \right] \cdot \mG(\vr'',\vr',\omega').
\end{align}
From Eqs.~(B1) and (B3) of Ref.~\onlinecite{suttorp04},
the first term is evaluated as
\begin{align}
& \left[ \cvJ(\vs'',\ii\omega+\delta),
  \dgg{\bvJ(\vr',-\ii\omega'+\delta) + \cvJ(\vr',\ii\omega'+\delta)} \right]
\nonumber \\ &
= \frac{2\hbar\diez\Im[\dieb(\vr',\omega')]\omega\omega'}
       {\ii(\omega-\omega'-\ii\delta)} \munit \delta(\vs''-\vr').
\end{align}
Since the second term obeys Eq.~\eqref{eq:G[J,Jd]G=(G-G)/(w-w)},
we obtain
\begin{align}
& \int\dd\vs''\ \mG^*(\vs',\vs'',\omega) \cdot
  \left[ \cvJ(\vs'',\ii\omega+\delta), \dgg{\hvJ(\vr',\omega')} \right]
\nonumber \\ &
= \frac{\diez\hbar}{\ii\pi}
  \frac{\Im[\dieb(\vr',\omega')]{\omega'}^3}
       {\omega(\omega-\omega'-\ii\delta)} \mG(\vs',\vr',\omega').
\end{align}
From Eqs.~(52) and (53a) of Ref.~\onlinecite{suttorp04},
the last term of Eq.~\eqref{eq:[cvPex,hvJd]} becomes
\begin{align}
& \int\dd\vs'\ \vdimP_{\mu'}^*(\vs') \cdot
  \left[ \ovA(\vs',0), \dgg{\hvJ(\vr',\omega')} \right]
\nonumber \\ &
= \frac{\hbar{\omega'}^2}{\pi c^2}\ \Im[\dieb(\vr',\omega')]
  \int\dd\vs'\ \trans{\vdimP_{\mu'}^*(\vs')} \cdot \mG(\vs',\vr',\omega').
\end{align}
Therefore, $\mX_1$ is evaluated as
\begin{align}
& \mX_1
= - \frac{\hbar\omega^2{\omega'}^2}{\pi^2c^4}\ \Im[\dieb(\vr,\omega)]\
  \Im[\dieb(\vr',\omega')]
\nonumber \\ & \times
 \sum_{\mu,\mu'} \int\dd\vs\
  \mG^*(\vr,\vs,\omega) \cdot \vdimP_{\mu}(\vs)\
  \crd_{\mu',\mu}^*(\omega)
\nonumber \\ & \times
  \int\dd\vs'\left[
    \frac{\omega\omega'\vdimP_{\mu'}^*(\vs')}{\omega-\omega'-\ii\delta}
     + \omega \trans{\vdimP_{\mu'}^*(\vs')} \right]
  \cdot \mG(\vs',\vr',\omega').
\label{eq:rep-mX_1} 
\end{align}
From its complex conjugate, we obtain
\begin{align}
& \mX_2 = - \frac{\hbar\omega^2{\omega'}^2}{\pi^2c^4}\ \Im[\dieb(\vr,\omega)]\
  \Im[\dieb(\vr',\omega')]
\nonumber \\ & \times
 \sum_{\mu,\mu'} \int\dd\vs\
  \mG^*(\vr,\vs,\omega) \cdot
  \left[ \frac{\omega\omega'\vdimP_{\mu}(\vs)}{\omega'-\omega+\ii\delta}
     + \omega' \trans{\vdimP_{\mu}(\vs)} \right]
\nonumber \\ & \times
  \crd_{\mu,\mu'}(\omega)
  \int\dd\vs'\ \vdimP_{\mu'}^*(\vs') \cdot \mG(\vs',\vr',\omega').
\end{align}
Since, from Eqs.~\eqref{eq:rwa-backward-P}, \eqref{eq:backward-src-ex}, 
\eqref{eq:b=Wsrc}, and \eqref{eq:[cex,cex]},
the commutator appearing in $\mX_3$ is written as
\begin{align}
& \biggl[ \cvPex(\vs,\ii\omega+\delta) + \frac{\ii}{\omega}\ovPexT(\vs,0),
\nonumber \\ & \quad
  \dgg{\cvPex(\vs',\ii\omega'+\delta)} - \frac{\ii}{\omega'}\ovPexT(\vs',0)
\biggr]
\nonumber \\ &
= \sum_{\mu,\mu'} \vdimP_{\mu}(\vs)\ \crd_{\mu',\mu}^*(\omega)
  \left[ \frac{\hbar\vdimP_{\mu'}^*(\vs')}{\omega-\omega'-\ii\delta}
       + \frac{\hbar}{\omega'} \trans{\vdimP_{\mu'}^*(\vs')} \right]
\nonumber \\ & \quad
- \sum_{\mu,\mu'}
  \left[ \frac{\hbar \vdimP_{\mu}(\vs)}{\omega-\omega'-\ii\delta}
       - \frac{\hbar}{\omega} \trans{\vdimP_{\mu}(\vs)} \right]
  \crd_{\mu,\mu'}(\omega')\ \vdimP_{\mu'}^*(\vs'),
\end{align}
we can find that
\begin{equation}
\mX_1 + \mX_2 + \mX_3 = \mzero.
\end{equation}
Therefore,
we obtain commutator \eqref{eq:[hvJz,hvJzd]} of the noise current density.
From the same kind of calculation and neglecting the nonresonance terms,
we also obtain Eq.~\eqref{eq:[hvJzd,hvJzd]}.

\subsection{\label{sec:with-abs}With nonradiative relaxation}
In this section, we perform the same kind of calculation 
as in the previous section
considering a nonradiative relaxation process of excitons.
Considering the reservoir oscillators,
the Laplace transforms of Heisenberg equation \eqref{eq:motion-ex-time} 
of excitons are derived as
\begin{subequations} \label{eq:motion-ex-L-damp} 
\begin{align}&
(\hbar\wex_{\mu}-\hbar\omega-\ii\delta)\ \bex_{\mu}(-\ii\omega+\delta)
\nonumber \\ &
= - \ii\hbar\oex_{\mu}(0) + \int\dd\vr\ \vdimP_{\mu}^*(\vr) \cdot
    \left[ \bvE(\vr,-\ii\omega+\delta) - \ovA(\vr,0) \right]
\nonumber \\ &
- \int_0^{\infty}\dd\wres
  \left[ \ccxr_{\mu}(\wres) \bres_{\mu}(\wres,-\ii\omega+\delta)
       + \ccxr^*_{\mu}(\wres) \bresd_{\mu}(\wres,-\ii\omega+\delta) \right]
\end{align}
\begin{align}&
(\hbar\wex_{\mu}-\hbar\omega+\ii\delta)\ \cex_{\mu}(\ii\omega+\delta)
\nonumber \\ &
= \ii\hbar\oex_{\mu}(0) + \int\dd\vr\ \vdimP_{\mu}^*(\vr) \cdot
    \left[ \cvE(\vr, \ii\omega+\delta) + \ovA(\vr,0) \right]
\nonumber \\ &
- \int_0^{\infty}\dd\wres
  \left[ \ccxr_{\mu}(\wres) \cres_{\mu}(\wres, \ii\omega+\delta)
       + \ccxr^*_{\mu}(\wres) \cresd_{\mu}(\wres, \ii\omega+\delta) \right].
\end{align}
\end{subequations}
On the other hand,
we obtain the motion equation of reservoir oscillators
\begin{equation} \label{eq:motion-res-time}
\ii\hbar\ddt{}\ores_{\mu}(\wres,t)
= \hbar\wres\ \ores_{\mu}(\wres,t) + \ccxr_{\mu}^*(\wres)
  \left[ \oex_{\mu}(t) + \oexd_{\mu}(t) \right],
\end{equation}
and its Laplace transforms
\begin{subequations} \label{eq:motion-res-L-damp} 
\begin{align}&
(\hbar\wres-\hbar\omega-\ii\delta) \bres_{\mu}(\wres,-\ii\omega+\delta)
\nonumber \\ &
= - \ii\hbar\ores_{\mu}(\wres,0) - \ccxr_{\mu}^*(\wres) \left[
    \bex_{\mu}(-\ii\omega+\delta) + \bexd_{\mu}(-\ii\omega+\delta) \right],
\end{align}
\begin{align}&
(\hbar\wres-\hbar\omega+\ii\delta) \cres_{\mu}(\wres, \ii\omega+\delta)
\nonumber \\ &
= \ii\hbar\ores_{\mu}(\wres,0) - \ccxr_{\mu}^*(\wres) \left[
    \cex_{\mu}(\ii\omega+\delta) + \cexd_{\mu}(\ii\omega+\delta) \right].
\end{align}
\end{subequations}
Substituting Eqs.~\eqref{eq:motion-res-L-damp} into \eqref{eq:motion-ex-L-damp}
and neglecting the nonresonant terms,
we obtain
\begin{subequations} \label{eq:motion-ex-L-damp2} 
\begin{align}&
\left[ \hbar\wex_{\mu}-\hbar\omega-\ii\damp_{\mu}(\omega)/2 \right]
  \bex_{\mu}(-\ii\omega+\delta)
\nonumber \\ &
= - \ii\hbar\oex_{\mu}(0) + \int\dd\vr\ \vdimP_{\mu}^*(\vr) \cdot
    \left[ \bvE(\vr,-\ii\omega+\delta) - \ovA(\vr,0) \right]
\nonumber \\ & \quad
+ \blg_{\mu}(-\ii\omega+\delta),
\end{align}
\begin{align}&
\left[ \hbar\wex_{\mu}-\hbar\omega+\ii\damp_{\mu}^*(\omega)/2 \right]
  \cex_{\mu}(\ii\omega+\delta)
\nonumber \\ &
= \ii\hbar\oex_{\mu}(0) + \int\dd\vr\ \vdimP_{\mu}^*(\vr) \cdot
    \left[ \cvE(\vr, \ii\omega+\delta) + \ovA(\vr,0) \right]
\nonumber \\ & \quad
+ \clg_{\mu}(\ii\omega+\delta),
\end{align}
\end{subequations}
where the operators on the RHS are defined as
\begin{subequations} \label{eq:def-blg-clg} 
\begin{align}
\blg_{\mu}(-\ii\omega+\delta)
& \equiv \int_0^{\infty}\dd\wres\
  \frac{\ii\ccxr_{\mu}(\wres)}{\wres-\omega-\ii\delta} \ores_{\mu}(\wres,0)
\nonumber \\ & \quad
- \int_0^{\infty}\dd\wres\
  \frac{\ii\ccxr_{\mu}^*(\wres)}{\wres+\omega+\ii\delta}
  \oresd_{\mu}(\wres,0),
\end{align}
\begin{align}
\clg_{\mu}(\ii\omega+\delta)
& \equiv - \int_0^{\infty}\dd\wres\
  \frac{\ii\ccxr_{\mu}(\wres)}{\wres-\omega+\ii\delta} \ores_{\mu}(\wres,0)
\nonumber \\ & \quad
+ \int_0^{\infty}\dd\wres\
  \frac{\ii\ccxr_{\mu}^*(\wres)}{\wres+\omega-\ii\delta} \oresd_{\mu}(\wres,0),
\end{align}
\end{subequations}
and the relaxation width is
\begin{equation} \label{eq:def-gamma} 
\frac{\ii\gamma_{\mu}(\omega)}{2} \equiv \int_0^{\infty}\dd\wres \left[
  \frac{|\ccxr_{\mu}(\wres)|^2}{\hbar\wres-\hbar\omega-\ii\delta}
+ \frac{|\ccxr_{\mu}(\wres)|^2}{\hbar\wres+\hbar\omega+\ii\delta}\right].
\end{equation}
Substituting Laplace transformed Maxwell wave equations
\eqref{eq:Laplace-Maxwell-3} into Eqs.~\eqref{eq:motion-ex-L-damp2}, 
the self-consistent equation sets
for the Laplace transformed exciton operators are obtained as
\begin{subequations} \label{eq:Laplace-self-consist-res} 
\begin{align}
&   \sum_{\mu'} \cssa_{\mu,\mu'}(\omega) \bex_{\mu'}(-\ii\omega+\delta)
\nonumber \\
& = \bsrc_{\mu}(-\ii\omega+\delta) + \blg_{\mu}(-\ii\omega+\delta)
  = \bsrca_{\mu}(-\ii\omega+\delta), \\
&   \sum_{\mu'} \{\cssa_{\mu',\mu}(\omega)\}^* \cex_{\mu'}(\ii\omega+\delta)
\nonumber \\
& = \csrc_{\mu}(\ii\omega+\delta) + \clg_{\mu}(\ii\omega+\delta)
  = \csrca_{\mu}(\ii\omega+\delta),
\label{eq:B-self-consist-res} 
\end{align}
\end{subequations}
and the one for the Fourier transform is
\begin{equation} \label{eq:self-consist-res} 
\sum_{\mu'} \cssa_{\mu, \mu'}(\omega)\ \hex_{\mu'}(\omega)
= \hsrc_{\mu}(\omega) + \hrsrc_{\mu}(\omega)
= \hsrca_{\mu}(\omega),
\end{equation}
where operator $\hrsrc_{\mu}(\omega)$ is defined as
\begin{align}\label{eq:def-hrsrc}
\hrsrc_{\mu}(\omega)
& \equiv \frac{1}{2\pi}
  \left[ \blg_{\mu}(-\ii\omega+\delta) + \clg_{\mu}(\ii\omega+\delta) \right]
\nonumber \\ & \quad
- \frac{1}{2\pi}
  \frac{\ii\damp_{\mu}(\omega) + \ii\damp_{\mu}^*(\omega)}{2}
  \cex_{\mu}(\ii\omega+\delta).
\end{align}
On the other hand, Fourier transformed motion equation
\eqref{eq:motion-hexone-damp} of the excitons is obtained 
by adding Eqs.~\eqref{eq:motion-ex-L-damp2}.
Further, by substituting Eq.~\eqref{eq:E=Ez+G*Pex} into it, 
we can directly obtain self-consistent equations 
\eqref{eq:self-consist-res}.

From commutation relations \eqref{eq:[ores,ores]} of $\ores_{\mu}(\wres)$,
the ones of the operators $\blg_{\mu}(-\ii\omega+\delta)$
and $\clg_{\mu}(\ii\omega+\delta)$ are calculated as
\begin{subequations} \label{eq:[lg,lg]} 
\begin{align}&
\left[ \blg_{\mu}(-\ii\omega+\delta),
  \dgg{\blg_{\mu'}(-\ii\omega'+\delta)}\right]
\nonumber \\ &
= \frac{\hbar\delta_{\mu, \mu'}}{\omega-\omega'+\ii\delta}
    \frac{\ii\damp_{\mu}(\omega) + \ii\damp_{\mu}^*(\omega')}{2},
\end{align}
\begin{align}&
\left[ \clg_{\mu}(\ii\omega+\delta),
  \dgg{\clg_{\mu'}(\ii\omega'+\delta)}\right]
\nonumber \\ &
= - \frac{\hbar\delta_{\mu, \mu'}}{\omega-\omega'-\ii\delta}
    \frac{\ii\damp_{\mu}^*(\omega) + \ii\damp_{\mu}(\omega')}{2},
\end{align}
\begin{align}&
\left[ \blg_{\mu}(-\ii\omega+\delta),
  \dgg{\clg_{\mu'}(\ii\omega'+\delta)}\right]
\nonumber \\ &
= - \frac{\hbar\delta_{\mu, \mu'}}{(\omega+\ii\delta)-(\omega'+\ii\delta)}
    \frac{\ii\damp_{\mu}(\omega) - \ii\damp_{\mu}(\omega')}{2},
\end{align}
and neglecting the nonresonant terms, we obtain
\begin{align}&
\left[ \blg_{\mu}(-\ii\omega+\delta), \blg_{\mu'}(-\ii\omega'+\delta) \right]
\nonumber \\ &
= \left[ \clg_{\mu}(\ii\omega+\delta), \clg_{\mu'}(\ii\omega'+\delta) \right]
\nonumber \\ &
= \left[ \blg_{\mu}(-\ii\omega+\delta), \clg_{\mu'}(\ii\omega'+\delta) \right]
= 0.
\end{align}
\end{subequations}
From these relations,
we obtian the commutation relations of $\csrca_{\mu}(\ii\omega+\delta)$:
\begin{align}
& \left[ \csrca_{\mu}(\ii\omega+\delta),
    \dgg{\csrca_{\mu'}(\ii\omega'+\delta)}\right] \nonumber \\
& = \hbar^2\delta_{\mu, \mu'}+ \frac{\hbar}{\omega-\omega'-\ii\delta}
    \left[ \rct_{\mu, \mu'}(\omega') - \rct_{\mu', \mu}^*(\omega) \right]
\nonumber \\ & \quad
  - \frac{\hbar\delta_{\mu, \mu'}}{\omega-\omega'-\ii\delta}
    \frac{\ii\damp_{\mu}^*(\omega) + \ii\damp_{\mu}(\omega')}{2} \\
& = \frac{\hbar}{\omega-\omega'-\ii\delta}
    \left[ \cssa_{\mu, \mu'}(\omega') - \{\cssa_{\mu', \mu}(\omega)\}^*\right],
\end{align}
and
\begin{equation}
\left[ \csrca_{\mu}(\ii\omega+\delta), \csrca_{\mu'}(\ii\omega'+\delta) \right]
= 0.
\end{equation}
They have a good correspondece with Eqs.~\eqref{eq:[src,srcd]=S-S} and
\eqref{eq:[src,src]=0}, which are discussed without the nonradiative 
relaxation.
Therefore, the commutation relations of $\cex_{\mu}(\ii\omega+\delta)$
have the same form as that of Eq.~\eqref{eq:[cex,cex]} except for replacing
$\crd_{\mu, \mu'}(\omega)$ with $\crda_{\mu, \mu'}(\omega)$,
and relations \eqref{eq:[hvJz,hvJz]} of $\hvJz(\vr, \omega)$ are not
changed even by considering the nonradiative relaxation.

Next, we evaluate the commutators of $\hrsrc_{\mu}(\omega)$
defined in Eq.~\eqref{eq:def-hrsrc}.
Here, we apply the Markov approximation to the reservoir oscillators
interacting  with excitons, 
and assume that correction term $\damp_{\mu}(\omega)$ is a real value.
From Eqs.~\eqref{eq:[lg,lg]} and Dirac's equivalence
$(x\pm\ii\delta)^{-1} = \text{P}x^{-1} \mp \ii\pi\delta(x)$, we find
\begin{subequations}
\begin{align}&
\left[ \blg_{\mu}(-\ii\omega+\delta) + \clg_{\mu}(\ii\omega+\delta),
\right. \nonumber \\ & \quad \left.
 \dgg{\blg_{\mu'}(-\ii\omega'+\delta) + \clg_{\mu'}(\ii\omega'+\delta)}\right]
\nonumber \\ &
= \delta_{\mu,\mu'} \delta(\omega-\omega') 2\pi\hbar \damp_{\mu}(\omega),
\end{align}
\begin{align}&
\left[ \blg_{\mu}(-\ii\omega+\delta) + \clg_{\mu}(\ii\omega+\delta),
\right. \nonumber \\ & \quad \left.
  \blg_{\mu'}(-\ii\omega'+\delta) + \clg_{\mu'}(\ii\omega'+\delta) \right]
= 0.
\end{align}
\end{subequations}
On the other hand, Eq.~\eqref{eq:B-self-consist-res} gives the relation
\begin{align}
&   \left[ \blg_{\mu}(-\ii\omega+\delta) + \clg_{\mu}(\ii\omega+\delta),
    \dgg{\cex_{\mu'}(\ii\omega'+\delta)}\right] \nonumber \\
& = - \biggl[
      \frac{\ii\hbar}{(\omega+\ii\delta)-(\omega'+\ii\delta)}
      \frac{\damp_{\mu}(\omega) - \damp_{\mu}(\omega')}{2}
\nonumber \\ & \quad
  + \frac{\ii\hbar}{\omega-\omega'-\ii\delta}
    \frac{\damp_{\mu}^*(\omega) + \damp_{\mu}(\omega')}{2}
  \biggr] \crda_{\mu,\mu'}(\omega').
\label{eq:[blg+clg,cexd]} 
\end{align}
Then, we obtain Eq.~\eqref{eq:[hrsrc,hrsrcd]}
and also Eq.~\eqref{eq:[hrsrcd,hrsrcd]} by neglecting the nonresonent terms.

Finally, we verify the independence of source operators 
$\hvJz(\vr,\omega)$ and $\hrsrc_{\mu}(\omega)$.
Since $\hrsrc_{\mu}(\omega)$ is defined as \eqref{eq:def-hrsrc},
we obtain the expression of the commutator
\begin{align}
&   \left[ \hrsrc_{\mu}(\omega), \dgg{\hvJz(\vr,\omega')} \right] \nonumber \\
& = \frac{1}{2\pi}
  \left[ \blg_{\mu}(-\ii\omega+\delta) + \clg_{\mu}(\ii\omega+\delta),
    \dgg{\hvJz(\vr,\omega')} \right] \nonumber \\ & \quad
- \frac{\ii}{2\pi}
  \frac{\damp_{\mu}(\omega) + \damp_{\mu}^*(\omega)}{2}
  \left[ \cex_{\mu}(\ii\omega+\delta), \dgg{\hvJz(\vr,\omega')} \right].
\label{eq:rep-[hrsrc,hvJzd]} 
\end{align}
Here, from Eqs.~\eqref{eq:rep-hvJz-P} and \eqref{eq:[blg+clg,cexd]},
the first term becomes
\begin{align}&
\frac{\ii\hbar{\omega'}^3}{4\pi^2c^2}\ \Im[\dieb(\vr,\omega')]
\nonumber \\ & \times
\left[
  \frac{\damp_{\mu}(\omega)-\damp_{\mu}(\omega')}
       {(\omega+\ii\delta)-(\omega'+\ii\delta)}
+ \frac{\damp_{\mu}^*(\omega)+\damp_{\mu}(\omega')}{\omega-\omega'-\ii\delta}
\right]
\nonumber \\ & \times
\sum_{\mu'} \crda_{\mu,\mu'}(\omega') \int\dd\vr'\ \vdimP_{\mu'}^*(\vr')
\cdot \mG(\vr',\vr,\omega').
\label{eq:rep-[hrsrc,hvJzd]-1st} 
\end{align}
The commutator appearing in the second term of
\eqref{eq:rep-[hrsrc,hvJzd]} is written as
\begin{align}&
\left[ \cex_{\mu}(\ii\omega+\delta), \dgg{\hvJz(\vr,\omega')} \right]
= \left[ \cex_{\mu}(\ii\omega+\delta), \dgg{\hvJ(\vr,\omega')} \right]
\nonumber \\ &
- \frac{{\omega'}^3}{\pi c^2}\ \Im[\dieb(\vr,\omega')]
  \int\dd\vr'\ \biggr[ \cex_{\mu}(\ii\omega+\delta), 
\nonumber \\ & \quad
  \dgg{\cvPex(\vr',\ii\omega'+\delta)}
       - \frac{\ii}{\omega'} \ovPexT(\vr',0) \biggl]
  \cdot \mG(\vr',\vr,\omega').
\label{eq:rep-[cex,hvJz]} 
\end{align}
From Eq.~\eqref{eq:rep-mX_1}, 
the first term on the RHS of \eqref{eq:rep-[cex,hvJz]}
\begin{align}&
\frac{\hbar{\omega'}^2}{\pi c^2}\ \Im[\dieb(\vr,\omega')] \sum_{\mu'}
  \cjg{\crda_{\mu',\mu}(\omega)}
\nonumber \\ & \times
  \int\dd\vr'
  \left[ \frac{\omega' \vdimP_{\mu'}^*(\vr')}{\omega-\omega'-\ii\delta}
       + \trans{\vdimP_{\mu'}^*(\vr')} \right] \cdot \mG(\vr',\vr,\omega'),
\end{align}
and from Eqs.~\eqref{eq:backward-src-ex}, \eqref{eq:B-self-consist-res}, and
\eqref{eq:[cex,cex]}, the second term is calculated as
\begin{align}&
- \frac{\hbar{\omega'}^2}{\pi c^2}\ \Im[\dieb(\vr,\omega')]
\nonumber \\ & \times
   \sum_{\mu'}
  \int\dd\vr'\ \biggl\{
    \frac{\omega' \vdimP_{\mu'}^*(\vr')}{\omega-\omega'-\ii\delta}
    \left[ \cjg{\crda_{\mu',\mu}(\omega)} - \crda_{\mu,\mu'}(\omega') \right]
\nonumber \\ & \quad
  + \cjg{\crda_{\mu',\mu}(\omega)} \trans{\vdimP_{\mu'}^*(\vr')}
  \biggr\} \cdot \mG(\vr',\vr,\omega').
\end{align}
Therefore, commutator \eqref{eq:rep-[cex,hvJz]} is evaluated as
\begin{align}&
\left[ \cex_{\mu}(\ii\omega+\delta), \dgg{\hvJz(\vr,\omega')} \right]
= \frac{\hbar{\omega'}^3}{\pi c^2}\
  \frac{\Im[\dieb(\vr,\omega')]}{\omega-\omega'-\ii\delta}
\nonumber \\ & \times
  \sum_{\mu'} \crda_{\mu,\mu'}(\omega')
  \int\dd\vr'\ \vdimP_{\mu'}^*(\vr') \cdot \mG(\vr',\vr,\omega').
\end{align}
From this and Eq.~\eqref{eq:rep-[hrsrc,hvJzd]-1st},
we obtain Eqs.~\eqref{eq:[hrsrc,hvJz]} by assuming 
$\damp_{\mu}(\omega) = \damp_{\mu}^*(\omega)$ 
and neglecting the nonresonant terms.
}
{
\def\TT{\mathrm{T}}

\def\RGA{D^{\mathrm{R}}}
\def\RGE{G^{\mathrm{R}}}
\def\RGX{D^{\mathrm{Rex}}}

\def\pv{\text{P}}

\def\ovPi{\mathbf{\Pi}}
\def\oPex{P_{\text{ex}}}
\def\dimP{\mathcal{P}}

\section{Self-standing modes and retarded correlation function\label{app:ssm}}
As discussed in sections \ref{sec:self-consist} and \ref{sec:absorption},
commutation relations of exciton operators 
in the Fourier transformed Heisenberg representation 
are written as Eq.~\eqref{eq:[hexone,hexone]}
in terms of inversed matrix $\mcrda(\omega)$ of coefficient
$\mcssa(\omega)$ of the self-consistent equation set.
As the analogue of the relation between the electric field and 
Green's tensor $\mG(\vr,\vr',\omega)$ satisfying 
Eq.~\eqref{eq:satisfied-mG},
we can understand that the elements of $\mcrda(\omega)$ identify
with the Fourier transforms of retarded correlation functions
$\RGX_{\mu,\mu'}$ of exciton operators:
\begin{equation}
- \hbar\ \crda_{\mu,\mu'}(\omega)
= \RGX_{\mu,\mu'}(\omega)
= \int_{-\infty}^{\infty}\dd t\ \RGX_{\mu,\mu'}(t)\
  \ee^{\ii\omega t},
\end{equation}
wehre 
\begin{equation} \label{eq:def-ret-G-ex} 
\RGX_{\mu,\mu'}(t-t') \equiv \begin{cases}
  - \ii \Braket{ \left[ \oex_{\mu}(t),  \oexd_{\mu'}(t') \right] }
& t > t' \\
  0 & t < t'
\end{cases}.
\end{equation}
This identity indicates that $\mcrda(\omega)$
has no pole in the upper half side of the complex $\omega$-plane.
On the other hand, 
the poles $\{\omega_{\lambda}\}$ in the lower half $\omega$-plane 
characterize the self-standing modes of the exciton-polaritons satisfying 
\begin{equation}
\det[\mcssa(\omega_{\lambda})] = 0,
\end{equation}
because it makes inverse matrix $\mcrda(\omega)$ singular 
at $\omega = \omega_{\lambda}$.

\section{Equal-time commutation relations\label{app:stcr}}
Commutation relations of equal-time Heisenberg operators
should keep the form of those of the Shr\"{o}dinger operators.
This means that the relations
\begin{equation} \label{eq:[oex(t),oexd(t)]} 
\left[ \oex_{\mu}(t), \oexd_{\mu'}(t) \right] = \delta_{\mu,\mu'},
\end{equation}
\begin{equation} \label{eq:[ovPex(t),ovPex(t)]} 
\left[ \ovPex(\vr,t), \ovPex(\vr',t) \right] = \mzero,
\end{equation}
\begin{equation} \label{eq:[ovE(t),ovE(t)]} 
\left[ \ovE(\vr,t), \ovE(\vr',t) \right] = \mzero
\end{equation}
should be derived from the commutation relations of the Fourier transformed
Heisenberg operators.
Moreover, from Eqs.~\eqref{eq:rep-ovE} and \eqref{eq:Heisen-A},
the electric field is represented as 
\begin{equation}
\diez\ovE(\vr,t) = - \ovPi(\vr,t) - \diez\grad\ophi(\vr,t),
\end{equation}
and since $\ovPi(\vr,t)$ and $\ovA(\vr,t)$ satisfy commutation relation
\eqref{eq:[ovA,ovPi]}, the relation
\begin{equation} \label{eq:[ovE(t),ovA(t)]} 
\left[ \diez\ovE(\vr,t), \ovA(\vr',t) \right]
= \ii\hbar\mdeltaT(\vr-\vr')
\end{equation}
should also be derived.
For local dielectric media, the same kind of calculation has been performed 
by Kn\"{o}ll, Scheel, and Welsch (KSW).\cite{knoll01}

From the time-representation of exciton operator
\begin{equation}
\oex_{\mu}(t)
= \int_{-\infty}^{\infty}\dd\omega\ \hex_{\mu}(\omega)\ \ee^{-\ii\omega t}
\end{equation}
and relations \eqref{eq:[hexone,hexone]-damp}
for $\omega$-representation,
the equal-time commutation relation is written as
\begin{equation} \label{eq:rep-[oex(t),oexd(t)]} 
\left[ \oex_{\mu}(t), \oexd_{\mu'}(t) \right]
= \frac{\hbar}{\ii2\pi} \int_{-\infty}^{\infty}\dd\omega
  \left[ \crda_{\mu,\mu'}(\omega) - \cjg{\crda_{\mu',\mu}(\omega)} \right].
\end{equation}
In the limit $|\omega|\rightarrow\infty$,
as indicated in App.~A.1 of KSW work,\cite{knoll01}
it is known that $\dieb(\vr,\omega)\rightarrow1$ and
\begin{equation} \label{eq:lim-w-infty-G} 
\lim_{|\omega|\rightarrow\infty} \frac{\omega^2}{c^2}\mG(\vr,\vr',\omega)
= - \delta(\vr-\vr').
\end{equation}
Then, due to the orthogonality of $\vdimP_{\mu}(\vr)$ shown 
in Eq.~\eqref{eq:rep-vdimP} and the relation with LT splitting
$\DLT^{\mu} = |\vdimP_{\mu}|^2/\dieb(\wex_{\mu})\diez$,
the limit of correction term \eqref{eq:def-rct} is
\begin{align}
\lim_{|\omega|\rightarrow\infty} \rct_{\mu,\mu'}(\omega)
& = \frac{1}{\diez} \int\dd\vr\ \vdimP_{\mu}^*(\vr) \cdot \vdimP_{\mu'}(\vr) \\
& = \delta_{\mu,\mu'}\ \dieb(\wex_{\mu})\DLT^{\mu}.
\end{align}
On the other hand, the limit of nonradiative width $\damp_{\mu}(\omega)$
defined in Eq.~\eqref{eq:def-gamma} is $\damp_{\mu}(\omega) \rightarrow 0$.
Therefore, coefficient matrix $\mcssa(\omega)$ becomes diagonal as
\begin{equation}
\lim_{|\omega|\rightarrow\infty} \cssa_{\mu,\mu'}(\omega)
= \left[ \hbar\wex_{\mu} + \dieb(\wex_{\mu})\DLT^{\mu}
       - \hbar\omega - \ii\delta \right] \delta_{\mu,\mu'},
\end{equation}
and of course its inverse matrix is also diagonal:
\begin{equation} \label{eq:lim-w-infty-crd} 
\lim_{|\omega|\rightarrow\infty} \crda_{\mu,\mu'}(\omega)
= \frac{\delta_{\mu,\mu'}}{\hbar\wex_{\mu} + \dieb(\wex_{\mu})\DLT^{\mu}
  - \hbar\omega - \ii\delta}.
\end{equation}
As mentioned in App.~\ref{app:ssm}, 
since $\mcrda(\omega)$ has no pole in the upper half $\omega$-plane,
the integration over the real axis is evaluated as
\begin{equation}
\int_{-\infty}^{\infty}\dd\omega\ \crda_{\mu,\mu'}(\omega)
= \frac{\ii\pi}{\hbar}\ \delta_{\mu,\mu'}.
\end{equation}
We can find that
this equation reproduces commutation relation \eqref{eq:[oex(t),oexd(t)]}
from Eq.~\eqref{eq:rep-[oex(t),oexd(t)]}.

Next, we verify commutation relation \eqref{eq:[ovPex(t),ovPex(t)]}
of excitonic polarization $\ovPex(\vr,t)$.
Although we approximate its positive-frequency Fourier component 
as Eq.~\eqref{eq:Pex=P*ex-rwa} in Sec.~\ref{sec:self-consist},
here we describe it without the RWA as
\begin{equation}
\hvPex^+(\vr,\omega)
= \sum_{\mu} \left[ \vdimP_{\mu}(\vr)\ \hex_{\mu}(\omega)
                  + \vdimP_{\mu}^*(\vr)\ \dgg{\hex_{\mu}(-\omega)} \right].
\label{eq:rep-hvPex-hex-hexd} 
\end{equation}
This representation keeps the following relation derived from 
the definition of Fourier transform Eq.~\eqref{eq:def-hvE}:
\begin{equation}
\hvPex^+(\vr,\omega) = \hvPex^-(\vr,-\omega) = \dgg{\hvPex^+(\vr,-\omega)}
\end{equation}
From this relation, the time-representation can be written as
\begin{align}
&   \ovPex(\vr,t) \nonumber \\
& = \int_0^{\infty}\dd\omega
    \left[ \hvPex^+(\vr,\omega)\ \ee^{-\ii\omega t}\
         + \hvPex^-(\vr,\omega)\ \ee^{\ii\omega t} \right] \\
& = \int_{-\infty}^{\infty}\dd\omega\ \hvPex^+(\vr,\omega)\ \ee^{-\ii\omega t}
  = \int_{-\infty}^{\infty}\dd\omega\ \hvPex^-(\vr,\omega)\ \ee^{\ii\omega t}.
\label{eq:rep-ovPex(t)-hvPex} 
\end{align}
Therefore, the equal-time commutator becomes
\begin{align}
&   \left[ \ovPex(\vr,t), \ovPex(\vr',t) \right] \nonumber \\
& = \int_{-\infty}^{\infty}\dd\omega \int_{-\infty}^{\infty}\dd\omega'\
    \ee^{-\ii\omega t}
    \left[ \hvPex^+(\vr,\omega), \hvPex^-(\vr',\omega') \right]
    \ee^{\ii\omega't} \nonumber \\
& = \sum_{\mu} \left[
      \vdimP_{\mu}(\vr) \vdimP_{\mu}^*(\vr')
    - \vdimP_{\mu}^*(\vr) \vdimP_{\mu}(\vr')
    \right].
\label{eq:rep-[ovPex(t),ovPex(t)]} 
\end{align}
We cannot obtain the second term from $\omega$-representation 
\eqref{eq:Pex=P*ex-rwa} with the RWA.
Equation \eqref{eq:rep-[ovPex(t),ovPex(t)]} is also obtained 
using Eqs.~\eqref{eq:rep-ovPex-oex} and \eqref{eq:[oex(t),oexd(t)]} directly.
From Eq.~\eqref{eq:rep-[ovPex(t),ovPex(t)]},
we can reproduce equal-time commutation relation 
\eqref{eq:[ovPex(t),ovPex(t)]}:
\begin{align}
&   \left[ \ovPex(\vr,t), \ovPex(\vr',t) \right]_{\xi,\xi'} \nonumber \\
& = \sum_{\mu} \left[
      \dimP_{\mu}^{\xi}(\vr) \cjg{\dimP_{\mu}^{\xi'}(\vr')}
    - \cc
    \right] \\
& = \sum_{\mu} \left[
      \braket{0|\oPex^{\xi}(\vr)|\mu} \braket{\mu|\oPex^{\xi'}(\vr')|0}
    - \cc
    \right] \\
& = \braket{0| \left[ \oPex^{\xi}(\vr), \oPex^{\xi'}(\vr') \right] |0}
  = 0,
\end{align}
where $\xi, \xi' = x, y, z$, $\ket{\mu} = \oexd_{\mu}\ket{0}$,
and $\ket{0}$ indicates the ground state of the medium.

Next, we verify relation \eqref{eq:[ovE(t),ovE(t)]}
of the electric field operator.
From the representation of the positive-frequency Fourier component of 
electric field \eqref{eq:E=Ez+G*Pex} and that of excitonic polarization
\eqref{eq:rep-hvPex-hex-hexd}, 
instead of Eq.~\eqref{eq:hvEone=hvEz+E*hexone},
we can also write the electric field operator without the RWA:
\begin{align}&
\hvE^+(\vr,\omega)
= \hvEz^+(\vr,\omega)
\nonumber \\ &
+ \sum_{\mu} [ \vdimE_{\mu}(\vr,\omega) \hex_{\mu}(\omega)
             - \vdimE_{\mu}^*(\vr,-\omega) \dgg{\hex_{\mu}(-\omega)} ],
\end{align}
where we use the relation derived from 
Eqs.~\eqref{eq:def-vdimE} and \eqref{eq:def-vdimF} with
$\mG(\vr,\vr',\omega) = \cjg{\mG(\vr,\vr',-\omega^*)}$:
\begin{align}
\vdimE_{\mu}(\vr,\omega) & = \vdimF_{\mu}^*(\vr,-\omega), \\
\vdimF_{\mu}(\vr,\omega) & = \vdimE_{\mu}^*(\vr,-\omega).
\end{align}
Since the time-representation of the electric field is also written 
in the same form as Eq.~\eqref{eq:rep-ovPex(t)-hvPex}, 
we can evaluate its equal-time commutator as
\begin{widetext}
\begin{align}&
\left[ \ovE(\vr,t), \ovE(\vr',t) \right]
= \int_{-\infty}^{\infty}\dd\omega \int_{-\infty}^{\infty}\dd\omega'\
  \ee^{-\ii(\omega-\omega')t}
  \left[ \hvEz^+(\vr,\omega), \hvEz^-(\vr',\omega') \right]
\nonumber \\ &
+ \frac{\hbar}{\ii2\pi} \sum_{\mu,\mu'} \int_{-\infty}^{\infty}\dd\omega\
  \Bigl[
    \vdimE_{\mu}(\vr,\omega)\ \crda_{\mu,\mu'}(\omega)\
    \vdimF_{\mu'}(\vr',\omega)
  - \vdimF_{\mu}^*(\vr,\omega) \cjg{\crda_{\mu',\mu}(\omega)}
    \vdimE_{\mu'}^*(\vr',\omega)
\nonumber \\ & \quad
  + \vdimE_{\mu}^*(\vr,-\omega) \cjg{\crda_{\mu,\mu'}(-\omega)}
    \vdimF_{\mu'}^*(\vr',-\omega)
  - \vdimF_{\mu}(\vr,-\omega)\ \crda_{\mu',\mu}(-\omega)\
    \vdimE_{\mu'}(\vr',-\omega)
  \Bigr].
\label{eq:rep-[ovE(t),ovE(t)]} 
\end{align}
\end{widetext}
The first term is the equal-time commutator 
between the electric fields in the background medium;
then, it becomes zero as indicated by SW
or as calculated by KSW.
Based on the fact that $\crda_{\mu,\mu'}(\omega)$ and 
$\mG(\vr,\vr',\omega)$ have no pole
in the upper half $\omega$-plane as discussed in App.~\ref{app:ssm}, 
we evaluate the second term using the residue theorem.
Since $\mG(\vr,\vr',\omega)$ becomes \eqref{eq:lim-w-infty-G}
in the limit of $|\omega|\rightarrow\infty$, 
from Eqs.~\eqref{eq:def-vdimE} and \eqref{eq:def-vdimF}, we can obtain
\begin{align}
\lim_{|\omega|\rightarrow\infty} \vdimE_{\mu}(\vr,\omega)
& = - \vdimP_{\mu}(\vr) / \diez, \\
\lim_{|\omega|\rightarrow\infty} \vdimF_{\mu}(\vr,\omega)
& = - \vdimP_{\mu}^*(\vr) / \diez.
\end{align}
From these and Eq.~\eqref{eq:lim-w-infty-crd},
the second term of Eq.~\eqref{eq:rep-[ovE(t),ovE(t)]} becomes
\begin{align}&
\frac{\hbar}{\ii2\pi} \sum_{\mu,\mu'} \int_{-\infty}^{\infty}\dd\omega\
    \vdimE_{\mu}(\vr,\omega)\ \crda_{\mu,\mu'}(\omega)\
    \vdimF_{\mu'}(\vr',\omega)
\nonumber \\ &
= \frac{1}{2{\diez}^2} \sum_{\mu} \vdimP_{\mu}(\vr)\vdimP_{\mu}^*(\vr').
\end{align}
Calculating the other terms in the same way, we obtain
\begin{align}&
\left[ \ovE(\vr,t), \ovE(\vr',t) \right]
\nonumber \\ &
= \frac{1}{{\diez}^2} \sum_{\mu}
  \left[ \vdimP_{\mu}(\vr)\vdimP_{\mu}^*(\vr')
       - \vdimP_{\mu}^*(\vr)\vdimP_{\mu}(\vr') \right]
= \mzero,
\end{align}
then, equal-time commutation relation \eqref{eq:[ovE(t),ovE(t)]} 
is reproduced.

Last, we verify relation \eqref{eq:[ovE(t),ovA(t)]}.
As discussed in App.~A.2 of KSW work,
we can write the time-representation of the vector potential as
\begin{align}
&   \ovA(\vr,t) \nonumber \\
& = \lim_{\epsilon\rightarrow0}\int_{\epsilon}^{\infty}\dd\omega
    \int\dd\vs
    \left[ \frac{\ee^{-\ii\omega t}}{\ii\omega} \hvE^+(\vs,\omega)
         + \Hc \right]
    \cdot \mdeltaT(\vs-\vr), \\
& = \pv \int_{-\infty}^{\infty}\dd\omega\ \frac{\ee^{-\ii\omega t}}{\ii\omega}
    \int\dd\vs\ \hvE^+(\vs,\omega) \cdot \mdeltaT(\vs-\vr), \\
& = - \pv \int_{-\infty}^{\infty}\dd\omega\ \frac{\ee^{\ii\omega t}}{\ii\omega}
    \int\dd\vs\ \hvE^-(\vs,\omega) \cdot \mdeltaT(\vs-\vr),
\end{align}
where $\pv$ indicates the principal value integration.
From this representation, the commutator between the electric field
and the vector potential becomes
\begin{widetext}
\begin{align}&
\left[ \diez \ovE(\vr,t), \ovA(\vr',t) \right]
= - \diez
  \int_{-\infty}^{\infty}\dd\omega\ \pv\int_{-\infty}^{\infty}\dd\omega'\
  \frac{\ee^{-\ii(\omega-\omega')t} }{\ii\omega'} \int\dd\vs
  \left[ \hvEz^+(\vr,\omega), \hvEz^-(\vs,\omega') \right]
  \cdot \mdeltaT(\vs-\vr')
\nonumber \\ &
+ \frac{\diez\hbar}{2\pi} \sum_{\mu,\mu'} \pv \int_{-\infty}^{\infty}
  \frac{\dd\omega}{\omega} \int\dd\vs\
  \Bigl[
    \vdimE_{\mu}(\vr,\omega)\ \crd_{\mu,\mu'}(\omega)\
    \vdimF_{\mu'}(\vs,\omega)
  - \vdimF_{\mu}^*(\vr,\omega)\ \crd_{\mu',\mu}^*(\omega)\
    \vdimE_{\mu'}^*(\vs,\omega)
\nonumber \\ & \quad
  + \vdimE_{\mu}^*(\vr,-\omega)\ \crd_{\mu,\mu'}^*(-\omega)\
    \vdimF_{\mu'}^*(\vs,-\omega)
  - \vdimF_{\mu}(\vr,-\omega)\ \crd_{\mu',\mu}(-\omega)\
    \vdimE_{\mu'}(\vs,-\omega)
  \Bigr] \cdot \mdeltaT(\vs-\vr').
\label{eq:rep-[eE(t),A(t)]} 
\end{align}
\end{widetext}
The first term is the same kind of commutator for background medium,
then it becomes $\ii\hbar\mdeltaT(\vr-\vr')$
as verified by KSW.
Therefore, all we have to do is verify that the other terms become zero.
First, in the limit of $|\omega|\rightarrow\infty$,
they becomes zero because of factor $\omega^{-1}$ 
comparing to the last four terms of Eq.~\eqref{eq:rep-[ovE(t),ovE(t)]}.
In the limit of $|\omega|\rightarrow0$, 
due to the equation shown in KSW work,
\begin{equation}
\lim_{|\omega|\rightarrow0} \frac{\omega^2}{c^2} \int\dd\vs\
  \mG(\vr,\vs,\omega) \cdot \mdeltaT(\vs-\vr') = \mzero
\end{equation}
we can find that the terms also become zero:
\begin{align}&
\lim_{|\omega|\rightarrow0} \frac{\omega^2}{c^2} \int\dd\vs\
  \mdeltaT(\vs-\vr') \cdot \vdimE_{\mu}(\vs,\omega)
\nonumber \\ &
= \lim_{|\omega|\rightarrow0} \frac{\omega^2}{c^2} \int\dd\vs\
  \mdeltaT(\vs-\vr') \cdot \vdimF_{\mu}(\vs,\omega)
= \mzero.
\end{align}
Therefore, Eq.~\eqref{eq:rep-[eE(t),A(t)]} 
reproduces equal-time commutation relation \eqref{eq:[ovE(t),ovA(t)]}.
}


\end{document}